\documentclass[aps,prd,twocolumn,groupedaddress,longbibliography,showpacs
,nofootinbib
]{revtex4-2}

\usepackage{amsmath,amssymb,latexsym,bm,hyperref}
\usepackage{graphicx,float}
\usepackage{xcolor}
\hypersetup{
    colorlinks,
    linkcolor={blue!80!black},
    citecolor={blue!80!black},
    urlcolor={blue!80!black}
}

\newcommand{\cross }[1]{#1 \hspace{-0.4em} /}

\begin{document}

\title{Clothed particle representation in quantum field theory: 
Fermion mass renormalization due to vector boson exchange}

\author{Yan Kostylenko}
\email[]{nlokost@gmail.com}
\author{Aleksandr Shebeko}
\email[]{shebekokipt@gmail.com}
\affiliation{Akhiezer Institute for Theoretical Physics, National Science Center "Kharkiv Institute of
Physics and Technology",  Kharkiv, Ukraine}

\date{\today}

\begin{abstract}

We consider the fermion mass renormalization due to the vector boson exchange within mesodynamics with nucleon and $\rho$ meson fields as well as quantum electrodynamics with electron and photon fields. 
The method of unitary clothing transformations is used to handle the so-called clothed particle representation that allows us to get rid of mass counterterms directly in the Hamiltonian. Thus, they can no longer appear in the $S$-matrix. 
Special attention is paid to the cancellation of the so-called contact terms that are inevitable in models with vector bosons.
Within this formalism, the second-order mass shifts are derived. They are expressed through the corresponding three-dimensional integrals whose integrands depend on certain covariant combinations of the relevant three-momenta. 
Our results are proved to be particle-momentum independent and compared with ones obtained by Feynman techniques.

\end{abstract}



\maketitle


\section{Introduction\label{intro}}

Starting from the instant form of relativistic quantum dynamics for a system of interacting particles, where amongst the ten generators of the Poincar\'e group $\Pi$ only the Hamiltonian $H$ and the boost operator $\mathbf{B}$ carry interactions, we have built up in the so-called clothed particle representation (CPR) the Hamiltonian for the interacting nucleon and $\rho$ meson fields, on the one hand, and for interacting electron and photon fields, on the other hand. In this connection, let as remind  that the transition from the primary "bare" particle representation (BPR) with its bare particle states to the CPR is implemented via the method of unitary clothing transformations (UCT) put forward by Greenberg and Schweber \cite{GreShv} and developed in Refs. \cite{ShiVis1974, SheShi1998, SheShi2000, SheShi2001, KorShe2004, KoCaSh2007, SheDub2010, FroShe2012, She16, Guilford, ArsEtAl2021, ArsKosSheBook} and \cite{Stef2001}. The corresponding clothing procedure is realized along the chain: bare particles with bare masses $\rightarrow$ bare particles with physical masses $\rightarrow$ physical (observable) particles (details in \cite{KoCaSh2007}). Such a consideration is convenient when drawing some parallels between the clothing approach in quantum field theory (QFT) and the method of canonical transformations (in particular, the mass-changing Bogoliubov-type ones) in the theory of superfluidity and superconductivity. 
As a result, all generators of $\Pi$ get one and the same sparse structure after normal ordering the creation (destruction) operators of clothed particles and removing the so-called bad terms (see survey \cite{SheShi2000}). Doing so, the state $\psi_{\mathbf{p}}$ of the clothed particle with energy $E_{\mathbf{p}}$ and momentum $\mathbf{p}$, being the total $H$ eigenvector, $H | \psi_{\mathbf{p}} \rangle=E_{\mathbf{p}}| \psi_{\mathbf{p}} \rangle$, belongs to the invariant subspace of the Fock space $R_F$ with mass $m$. The latter is determined by relation
\begin{equation}\label{masses_def}
    m^2=E_{\mathbf{p}}^2-\mathbf{p}^2
\end{equation}
while in the contemporary renormalization theory the particle mass is considered as a pole in the exact particle propagator (see, e.g., 
Chapters 10, 11 in \cite{Wein1}).
Unlike this we prefer to use the natural definition \eqref{masses_def} so 
\begin{equation}\label{mass_def}
    H | \psi_{\mathbf{p}=0 } \rangle = m| \psi_{\mathbf{p}=0 }\rangle 
\end{equation}
and introduce the mass shift $\delta m = m - m_0$, where $m_0$ is the bare mass value (e.g., for a fermion) with inequality
\begin{equation}
    H | \psi^{\textrm{bare}}_{\mathbf{p}=0 } \rangle \neq m_0| \psi^{\textrm{bare}}_{\mathbf{p}=0 }\rangle ,
\end{equation}
for the one-body bare state $| \psi^{\textrm{bare}}_{\mathbf{p}=0} \rangle \neq | \psi_{\mathbf{p}=0 } \rangle$.

Recall also that within the  Dyson-Feynman approach or old-fashioned perturbation theory the mass shifts may be expressed through the corresponding  self-energy contributions to the $S$-matrix. Such contributions enter an expansion of the mass shifts in the coupling constants
and some of them give rise to undesirable divergences inherent in the existing applications of every local field model (at least, when employing the perturbative methods). 
Their removal requires considerable efforts associated with a consequent regularization of the divergent integrals involved. 
In the S-matrix calculations they are encountered to the first non-vanishing approximation in the coupling constants (in particular, when evaluating the forward scattering amplitudes, where the one-loop contributions must cancel the occurring mass counterterms). A few instructive examples of such a situation for the pion-nucleon and nucleon-nucleon scattering amplitudes can be found in the monograph \cite{Schweber} both in the framework of the old-fashioned perturbation theory and the Dyson-Feynman approach. 

One should note that the clothed particle approach used here is related to an extensive topic associated with the renormalization and dressing procedures in the QFT. First, we mean pioneering explorations in the framework of the so-called similarity renormalization group (SRG) method put forward by Głazek and Wilson (see, e.g., Ref. \cite{GlaWil1993} and Refs. therein). Second, we have seen the promising applications of the akin approach proposed by Wegner \cite{Weg1994} with its flow equations for Hamiltonians. Both approaches are aimed at softening interactions in nuclear systems, if any, in order to improve the convergence of the corresponding computations. In this context, we would like to mention a few instructive applications \cite{Gla2012, Gla2013,AndBogFurPer2010} of the SRG evolved procedure with the simple model for primary interactions. Although our approach and those are implemented via a similarity transformation $O \to U O U^{-1}$ ($U^{-1}=U^\dag$) of any operator $O$, the objective of each of such a transformation is completely different.
Moreover, by definition, the UCTs remain the primary Hamiltonian intact. 
Being pressed in space, we can not discuss all the attractive features of these approaches. Certainly, we are going to do it later.

This work goes on our studies \cite{KorShe2004}.
In addition, we would like to mention work \cite{KruGlo1999}, 
where we have seen some points similar to ours.
The authors of Ref. \cite{KruGlo1999}  have chosen the definition (\ref{masses_def}) to calculate the second-order mass shift of scalar "nucleon" due to exchange of scalar meson by using the 
so-called FST - Okubo method \cite{FST, Okubo}.
They have proved that their result is equivalent to the one obtained via the Dyson-Feynman techniques. 
The approach exposed here differs from the aforementioned one, at least, in the two aspects. First, unlike the approach \cite{KruGlo1999} the clothing procedure
is not aimed, \textit{a priori}, to find a UT that blockdiagonalizes $H$. At this point one should note that such a reduction of the Hermitian operator via a unitary transformation can not be implemented, in general, in infinite-dimensional Fock space.
Instead, the aim of the multistep clothing procedure is to express the original Hamiltonian $H$ in terms of the new clothed-particle operators in a form which is different from that given for the initial bare-particle one. Such a transition from the BPR to the CPR introduces a new sparse structure of the original Hamiltonian $H$. Second, in the framework of the CPR the mass and vertex renormalization problem \cite{SheShi2001, KorShe2004} is considered in a natural way, in parallel with the construction of the interactions.

Since the publication of \cite{KorShe2004}, a number of applications of the UCT method have been notably extended. Currently, it has been applied in the mesodynamics to the processes that involve interactions of the clothed $\pi$, $\eta$, $\rho$, $\omega$, $\delta$, $\sigma$ mesons and nucleons \cite{SheShi2001, KoCaSh2007, SheDub2010, DubShe13, She13, KamSheArs17}. The applications of this method in QED and QCD have been shown in \cite{She14, She16, Caen, Guilford, ArsKosShe2021}. In the present work we return to the mass renormalization problem in the CPR to consider fermion mass renormalization due to the vector bosons exchange.

The outline of this paper is as follows.
In Sec. \ref{formalism}, we will take a look briefly at the underlying formalism and structure of the Hamiltonian.
A distinctive feature of the vector coupling is that the corresponding Lorentz-invariant Lagrangian does not necessarily have ``\textit{... the interaction Hamiltonian as the integral over space of a scalar interaction density: we also need to add non-scalar terms to the interaction density ...}'' (quoted from p. 292 of Ref. \cite{Wein1}). It is the case with derivative couplings and/or spin $\geq 1$, see Eqs. (77) and (81-83) in \cite{SheDub2010} (note a typo in Eqs. (83, 91): instead of $4m^2$, the denominator of the second term should be $8m^2$). As shown in \cite{SheDub2010}, the clothing procedure enables us to remove the non-invariant terms which belong to $NN \rightarrow NN$ interaction directly in the Hamiltonian (at least, in the second order in coupling constants). This pleasant feature of the CPR works well in the present problem, viz., non-invariant mass renormalization terms cancel too.
In Sec. \ref{mass_shifts_UCT}, we will derive the mass shifts of electron and nucleon due to photon and $\rho$ meson exchanges, respectively. This approach can be easily carried over to the cases of exchanges with other vector particles, for example, the $\omega$ meson.
We will also consider a way to regularize the integrals included in the expressions for the mass shifts in Sec. \ref{Cutoff}.
Our results will be compared with the simplest disconnected contributions to the corresponding boson-fermion forward scattering amplitude in Sec. \ref{comparison}. They are evidently covariant and determined by the one-loop diagrams.
Sec.~\ref{comparison_with_other} has been written to understand better the place of the notion of clothed particles in the framework of a popular version of the $in(out)$ formalism.
In particular, there we find the touching points between the CPR approach and Lehmann–Symanzik–Zimmermann (LSZ) method in QFT.
At last, the Appendix \ref{KNN_derivation} is devoted to the derivation of the $2 \leftrightarrow 2$ interactions between the clothed nucleons (antinucleons).

\section{Underlying formalism\label{formalism}}

Here the notion of clothed particles is applied to the following model: a spinor (fermion) field $\psi$ interacts with a vector boson field $\varphi$. The model Hamiltonian can be separated into the free and interaction parts $H(\alpha)={H_F(\alpha)+H_I(\alpha)}$, where the interaction part 
\begin{equation}\label{HI_def}
    H_I(\alpha)=M_\textrm{ren}+V_\textrm{ren}+V^{(1)}+V^{(2)}
\end{equation} 
includes the mass $M_\textrm{ren}$ and vertex $V_\textrm{ren}$ counterterms. The interaction is a function of creation (destruction) operators in the BPR, i.e., referred to the bare particles with physical masses \cite{KoCaSh2007}. 
Formally, the latter can be introduced via the mass-changing Bogoliubov-type transformation.
We will use the well-known Yukawa-type interactions between the nucleon and $\rho$ meson fields, namely
\begin{multline}\label{V1_def}
    V^{(1)}=\int d \mathbf{x} 
    \Big(
        g\, :\bar\psi(\mathbf{x}) \gamma_\alpha \psi(\mathbf{x}): \varphi^\alpha(\mathbf{x})
        \\
        +\frac{f}{4m}\, :\bar\psi(\mathbf{x}) \sigma_{\alpha\beta} \psi(\mathbf{x}): \varphi^{\alpha\beta}(\mathbf{x})
    \Big)\,,
\end{multline}
\begin{multline}\label{V2_def}
    V^{(2)}=\int d \mathbf{x} 
    \Big(
        \frac{g^2}{2m_b^2}\, :\bar\psi(\mathbf{x}) \gamma_0 \psi(\mathbf{x}): :\bar\psi(\mathbf{x}) \gamma_0 \psi(\mathbf{x}):
        \\
        +\frac{f^2}{8m^2}\, :\bar\psi(\mathbf{x}) \sigma_{0i} \psi(\mathbf{x}): :\bar\psi(\mathbf{x}) \sigma_{0i} \psi(\mathbf{x}):
    \Big)\,,
\end{multline}
where $\sigma_{\alpha\beta}=\frac{i}{2}(\gamma_\alpha \gamma_\beta - \gamma_\beta \gamma_\alpha)$, $\varphi^{\alpha\beta}(\mathbf{x})=\partial^\alpha\varphi^\beta(\mathbf{x}) - \partial^\beta\varphi^\alpha(\mathbf{x})$, and symbol $:$ denotes the normal ordering, where all the creation operators are to the left of the destruction ones.
In the case of QED, we will work in the Coulomb gauge (see, for example, Eq. (8.4.23) in Ref. \cite{Wein1}), therefore we should take $g=e$, $f=0$ in (\ref{V1_def}) and the Coulomb interaction
\begin{multline}\label{VCoul_def}
    V^{(2)}=e^2 \int d \mathbf{x}\, d \mathbf{y}
    \frac{ e^{-\lambda |\mathbf{x}-\mathbf{y}|}}{4\pi|\mathbf{x}-\mathbf{y}|}
    \\\times\, :\bar\psi(\mathbf{x}) \gamma_0 \psi(\mathbf{x}): :\bar\psi(\mathbf{y}) \gamma_0 \psi(\mathbf{y}):
    \equiv V_\textrm{Coul}\,,
\end{multline}
with the parameter $\lambda$ to be put zero at the end of calculations.
Henceforth, the upper indices denote the order in coupling constants.
The set $\alpha$ involves the creation (destruction) operators for the bare bosons $a^\dagger(a)$, fermions $b^\dagger(b)$, and antifermions $d^\dagger(d)$. Following a common practice, they appear in the standard Fourier expansions of the boson and fermion fields
\begin{multline}\label{varphi_expan}
    \varphi_\alpha(\mathbf{x})=
    \frac{1}{\sqrt{2(2\pi)^3}}\int\frac{d \mathbf{k}}{\omega_{\mathbf{k}}}
    \big(
    e_\alpha(k\sigma)a(k\sigma) 
    \\
    +e_\alpha(k_-\sigma)a^\dagger(k_-\sigma)   
    \big)
    e^{i \mathbf{k}\cdot \mathbf{x}}\,,
\end{multline}
\begin{multline}\label{psi_expan}
    \psi(\mathbf{x})=
    \sqrt\frac{m}{(2\pi)^3}\int\frac{d \mathbf{p}}{E_{\mathbf{p}}}\big(
    u(p\mu)b(p\mu) 
    \\
    +\upsilon(p_-\mu)d^\dagger(p_-\mu)   
    \big)
    e^{i \mathbf{p}\cdot \mathbf{x}}\,,
\end{multline}
where $\omega_{\mathbf{k}}=\sqrt{m_b^2+\mathbf{k}^2}$, $E_{\mathbf{p}}=\sqrt{m^2+\mathbf{p}^2}$, $m_b$ and $m$ are boson ($\rho$ meson) and fermion (nucleon) masses, respectively. Hereafter, for brevity, we mean summation over dummy polarization indices ($\mu$, $\sigma$, etc.). The independent polarization vectors $e(k\sigma)$ ($\sigma=+1,0,-1$) in the last expansion are transverse $k^\alpha e_\alpha(k\sigma)=0$ and normalized as 
\begin{equation}\label{Pk_nucl}
    \sum_\sigma e_\alpha(k\sigma)e_\beta^*(k\sigma) =-g_{\alpha\beta}+\frac{k_\alpha k_\beta}{m_b^2} 
    \equiv P_{\alpha\beta}(k)\,.
\end{equation}
As usually, we apply the following commutation rules
$[a(k\sigma),a^\dagger(k'\sigma')]=\omega_{\mathbf{k}}\delta_{\sigma'\sigma}\delta(\mathbf{k}'-\mathbf{k})$,
$\{b(p\mu),b^\dagger(p'\mu')\}=\{d(p\mu),d^\dagger(p'\mu')\}=E_{\mathbf{p}}\delta_{\mu'\mu}\delta(\mathbf{p}'-\mathbf{p})$. 

In any case, we have the free part of $H(\alpha)$
\begin{multline}\label{HF_def}
    H_F(\alpha)=\int\frac{d \mathbf{p}}{E_{\mathbf{p}}} E_{\mathbf{p}} 
    [b^\dagger(p\mu) b(p\mu)+d^\dagger(p\mu) d(p\mu)]\\
    +\int\frac{d \mathbf{k}}{\omega_{\mathbf{k}}} \omega_{\mathbf{k}} 
    a^\dagger(k\sigma) a(k\sigma)
    \,.
\end{multline}

Explicitly non-covariant expressions (\ref{V2_def}) and (\ref{VCoul_def}) reflect the mentioned peculiarity of the interaction density in such models. Namely, it is not a Lorentz scalar. In this connection, let us recall the property of the density $V_D(x)=e^{iH_F t} V(\mathbf{x}) e^{-iH_F t}$ in the Dirac picture (D-picture) to be a scalar, viz.,
\begin{equation}
    U_F(\Lambda)V_D(x)U_F^{-1}(\Lambda) = V_D(\Lambda x)\,,
\end{equation}
where the operators $U_F(\Lambda)$ realize a unitary irreducible representation of the Poincar\'e group in the Hilbert space of states for free (non-interacting) fields.

For a set of clothed operators we will use the symbol $\alpha_c$. To apply the clothing procedure exposed in \cite{SheShi2001, KoCaSh2007} one has to proceed to the CPR via elimination of the so-called bad terms, which, by definition, prevent the bare vacuum and the bare one-particle states to be $H$ eigenvectors.
Thus, we will use the first clothing transformation $W^{(1)}(\alpha){=W^{(1)}(\alpha_c)}{=\exp(R^{(1)})}$ in order to eliminate the interaction $V^{(1)}$ linear in coupling constants.
It can be done if the generator $R^{(1)}$ meets the equation
\begin{equation} \label{R_def}
    \left[R^{(1)},H_F\right]+V^{(1)}=0.
\end{equation}
In the CPR with the similarity transformation ${\alpha = W(\alpha_c) \alpha_c W^\dagger(\alpha_c)}$ we have 
\begin{multline}
    H(\alpha) \equiv K(\alpha_c) = W(\alpha_c)H(\alpha_c)W^\dagger(\alpha_c) \\
    = e^R[H_F(\alpha_c)+H_I(\alpha_c)]e^{-R}\,.
\end{multline}
By using the definition (\ref{R_def}) it can be written as
\begin{multline}\label{K_series}
    K(\alpha_c) =H_F +M_{ren} +V_{ren} +V^{(2)} +\frac{1}{2}[R^{(1)},V^{(1)}] \\
    +[ R^{(1)},M_{ren}] +\frac{1}{3}[R^{(1)},[R^{(1)},V^{(1)}]] +...\\
    =H_F(\alpha_c)+K_I(\alpha_c)\,.
\end{multline}
Keeping in the r.h.s. of \eqref{K_series} only the contributions of the second order in coupling constants we get
\begin{equation}\label{KI2}
    K_I(\alpha_c) \approx K^{(2)}_I(\alpha_c)=M^{(2)}_\textrm{ren} + V^{(2)} + \frac{1}{2}\big[R^{(1)},V^{(1)}\big]\,.
\end{equation}
In other words, we neglect terms responsible for the processes more complicated than $2\leftrightarrow 2$.
All particle-conserving ingredients of $M^{(2)}_\textrm{ren}$  should be cancelled by the corresponding one-body (e.g. $b_c^\dag b_c$) terms $K^{(2)}_\textrm{1-body}$ that involve components of the commutator $\frac{1}{2}\big[R^{(1)},V^{(1)}\big]$ and the operator $V^{(2)}$, that is done in the next section. 
As shown in \cite{KorShe2004}, only the particle-conserving part of the mass counterterm (responsible for the one to one fermion transition) may be cancelled via one and the same clothing transformation. 
Note that it is sufficient to evaluate the mass shifts in the second order since the same operator structure will appear in higher orders in the coupling constant. 
The operator $K^{(2)}_I$ involves the two-body (e.g. $b_c^\dag b_c^\dag b_c b_c$) interactions responsible for the physical processes between the clothed particles
\begin{multline}\label{KI}
    K^{(2)}_I=K(ff \rightarrow ff) + K(\bar{f}\bar{f} \rightarrow \bar{f}\bar{f}) 
    +
    K(f\bar{f} \rightarrow f\bar{f}) \\+ K(bf \rightarrow bf) + K(b\bar{f} \rightarrow b\bar{f}) 
    +
    K(bb \rightarrow f\bar{f}) + K(f\bar{f} \rightarrow bb)
    \\+
    K^{(2)}_\textrm{1-body}
    + M^{(2)}_\textrm{ren}.
\end{multline}
In Ref. \cite{ArsKosShe2021} we have presented all the four-operator interactions in QED, while in the Appendix \ref{KNN_derivation}  (cf. Ref. \cite{SheDub2010}) there are a step-by-step derivation of these four-operators interactions, that belong to the nucleon-antinucleon subsector in the mesodynamics with the vector coupling. One of these interactions is responsible for the nucleon-nucleon scattering
\begin{multline}\label{KNN}
    K(NN \rightarrow NN)=
    \frac{m^2}{2(2\pi)^3}\int 
    \frac{d\mathbf{p}'_1 d\mathbf{p}'_2 d\mathbf{p}_1 d\mathbf{p}_2}{E_{\mathbf{p}'_1} E_{\mathbf{p}'_2} E_{\mathbf{p}_1} E_{\mathbf{p}_2}}
    \\ \times
    \frac{\delta(\mathbf{p}'_1 + \mathbf{p}'_2 - \mathbf{p}_1 - \mathbf{p}_2)}{(p'_1-p_1)^2-m_b^2}
    b_c^\dagger(p'_1\mu'_1) b_c^\dagger(p'_2\mu'_2) b_c(p_1\mu_1) b_c(p_2\mu_2)
    \\ \times
    \bigg[
        \bar{u}(p'_1\mu'_1) \bigg\{ (g+f)\gamma_\alpha - \frac{f}{2m}(p'_1+p_1)_\alpha \bigg\} u(p_1\mu_1)
        \\ \times
        \bar{u}(p'_2\mu'_2) \bigg\{ (g+f)\gamma^\alpha - \frac{f}{2m}(p'_2+p_2)^\alpha \bigg\} u(p_2\mu_2) 
        \\
        + (E_{\mathbf{p}'_1} + E_{\mathbf{p}'_2} - E_{\mathbf{p}_1} - E_{\mathbf{p}_2})
        \frac{f}{2m} \bar{u}(p'_2\mu'_2) \gamma^0\bm{\gamma} u(p_2\mu_2)
        \\ \times
        \bar{u}(p'_1\mu'_1) \bigg\{ (g+f)\bm{\gamma} - \frac{f}{2m}(\mathbf{p}'_1+\mathbf{p}_1) \bigg\} u(p_1\mu_1)
    \bigg].
\end{multline}
Henceforth, we omit the index $c$ if it does not lead to confusion.
It is proved (see the Appendix \ref{KNN_derivation}) that the so-called contact interaction ${K_\textrm{cont}(NN \rightarrow NN)}$
cancels the contribution $V^{(2)}(NN \rightarrow NN)$ that stem from the non-scalar part of the interaction density $V^{(2)}(x)$, so
\begin{multline}
    \frac{1}{2}\big[R^{(1)},V^{(1)}\big](NN \rightarrow NN) \\
    = K(NN \rightarrow NN) + K_\textrm{cont}(NN \rightarrow NN)\,.
\end{multline}
It gives the opportunity to handle the interaction \eqref{KNN} that is covariant on the energy shell. 

\section{Derivation of the mass shifts\label{mass_shifts_UCT}}

According to \cite{SheShi2001} the commutator ${\big[R^{(1)},V^{(1)}\big]}=
{\big[ \mathcal{R}^{(1)} , \mathcal{V}^{(1)} \big]} + {\big[ \mathcal{R}^{(1)} , \mathcal{V}^{(1) \dagger} \big]} + \textrm{H.c.}$ with 
\begin{equation}\label{V1_and_matrix}
    \begin{split}
        &V^{(1)}=\mathcal{V}^{(1)} + \mathcal{V}^{(1) \dagger},
        \\
        &\mathcal{V}^{(1)}= \frac{m}{\sqrt{2(2\pi)^3}}\int\frac{d \mathbf{k}}{\omega_{\mathbf{k}}} \frac{d \mathbf{p}_1}{E_{\mathbf{p}_1}} \frac{d \mathbf{p}_2}{E_{\mathbf{p}_2}}
        \\
        &\times F^\dagger(p_1\mu_1) V(p_1\mu_1;p_2\mu_2;k\sigma) F(p_2\mu_2) a(k\sigma)\,,
    \end{split}
\end{equation}
\begin{equation}\label{R1_and_matrix}
    \begin{split}
        &R^{(1)}=\mathcal{R}^{(1)} - \mathcal{R}^{(1) \dagger},
        \\
        &\mathcal{R}^{(1)}= \frac{m}{\sqrt{2(2\pi)^3}}\int\frac{d \mathbf{k}}{\omega_{\mathbf{k}}} \frac{d \mathbf{p}_1}{E_{\mathbf{p}_1}} \frac{d \mathbf{p}_2}{E_{\mathbf{p}_2}}
        \\
        &\times F^\dagger(p_1\mu_1) R(p_1\mu_1;p_2\mu_2;k\sigma) F(p_2\mu_2) a(k\sigma)\,.
    \end{split}
\end{equation}
Here, the fermion operator column $F$ and row $F^\dagger$ are composed of the particle and antiparticle operators (e.g., $F^\dagger(p\mu)=[b^\dagger(p\mu),d(p_-\mu)]$), the c-number matrices
\begin{multline*}\label{Vij_def}
    V(p_1\mu_1;p_2\mu_2;k\sigma)= 
    \begin{bmatrix}
        V_{11} & V_{12}\\ 
        V_{21} & V_{22}
    \end{bmatrix}
    =
    \delta(\mathbf{p}_1-\mathbf{p}_2-\mathbf{k})
    e^\alpha(k\sigma)
    \\ \times
    \begin{bmatrix}
        \bar{u}(p_1\mu_1) \\
        \bar{\upsilon}(p_{1-}\mu_1)
    \end{bmatrix}
    (g\gamma_\alpha
    -\frac{f}{2m}ik^\beta\sigma_{\beta\alpha})
    \begin{bmatrix}
        u(p_2\mu_2) & \upsilon(p_{2-}\mu_2)
    \end{bmatrix}\,
\end{multline*}
and
\begin{equation}
    R(p_1\mu_1;p_2\mu_2;k\sigma)= 
    \begin{bmatrix}
        R_{11} & R_{12}\\ 
        R_{21} & R_{22}
    \end{bmatrix}\,,\\
\end{equation}
\begin{equation}
    R_{\varepsilon' \varepsilon}=\frac{V_{\varepsilon' \varepsilon}}{(-1)^{\varepsilon'-1}E_{\mathbf{p}_1}-(-1)^{\varepsilon-1}E_{\mathbf{p}_2}-\omega_{\mathbf{k}}}\,.
\end{equation}
At this point we recall that such a relation is valid if ${m_b<2m}$ (see Sec. 2.4 in \cite{SheShi2001}).

As mentioned above, all particle-conserving mass renormalization counterterms in 
$M^{(2)}_\textrm{ren} = {M^{(2)}_\textrm{bos} + M^{(2)}_\textrm{ferm}}$,
\begin{multline}\label{Mfermion_def}
    M^{(2)}_\textrm{ferm} = \delta m^{(2)} \int d \mathbf{x} \bar\psi(\mathbf{x})\psi(\mathbf{x})
    \\
    =m \,\delta m^{(2)} \int \frac{d\mathbf{p}}{E_\mathbf{p}^2}
    F^\dagger(p\mu_1) M(p\mu_1\mu_2) F(p\mu_2),
\end{multline}
\begin{equation}\label{Mmatrix}
    M(p\mu_1\mu_2)= 
    \begin{bmatrix}
        \delta_{\mu_1\mu_2} & 
        \bar u(p\mu_1) \upsilon(p_-\mu_2)\\ 
        \bar \upsilon(p_-\mu_1) u(p\mu_2) & 
        -\delta_{\mu_1\mu_2}
    \end{bmatrix},
\end{equation}
where $\delta m^{(2)}$ is the fermion mass shift in question,
should be cancelled by the corresponding contributions from the commutator $\frac{1}{2}\big[R^{(1)},V^{(1)}\big]$ and the operator $V^{(2)}$. 
It gives us the relationship
\begin{equation}\label{ren_requirement}
    M^{(2)}_\textrm{ren}\,_{b^\dagger b} + V^{(2)}_{b^\dagger b} + \frac{1}{2}\big[R^{(1)},V^{(1)}\big]_{b^\dagger b}=0.
\end{equation}
Subscript $b^\dagger b$ means that after normal ordering we retain only one-fermion terms. 
Note that the similar equation for one-antifermion terms ($d^\dagger d$) leads to the same results for antifermion mass shifts.
It turns out that non-diagonal ``bad'' $b^\dagger d^\dagger$ and $b d$ terms could not be cancelled with the respective terms of the operator $M^{(2)}_\textrm{ren}$, at least in the local theory.
In this context, let us write down all the one-fermion terms included in Eq. \eqref{ren_requirement}
\begin{equation}\label{Mren_bb}
    M^{(2)}_\textrm{ren}\,_{b^\dagger b}
    =
    m\,\delta m^{(2)} \int \frac{d \mathbf{p}}{E_{\mathbf{p}}^2}
    b^\dagger(p\rho)b(p\rho)\,,
\end{equation}
\begin{widetext}
\begin{multline}\label{V2}
    V^{(2)}_{b^\dagger b} = 
    \frac{m^2}{4(2\pi)^3}
    \int \frac{d \mathbf{p}}{E_{\mathbf{p}}^2}
    \frac{d \mathbf{q}}{E_{\mathbf{q}}\omega_{\mathbf{p}-\mathbf{q}}}
    b^\dagger(p\rho)b(p\rho) 
    \bar{u}(p\mu)\Bigg\{
    g^2 \frac{\omega_{\mathbf{p}-\mathbf{q}}}{m_b^2}
    \bigg[
        \gamma_0u(q\nu) \bar{u}(q\nu)\gamma_0
        -
        \gamma_0\upsilon(q_-\nu) \bar{\upsilon}(q_-\nu)\gamma_0
    \bigg]
    \\-
    f^2 \frac{\omega_{\mathbf{p}-\mathbf{q}}}{4m^2} 
    \bigg[
        \gamma_0\bm{\gamma}u(q\nu) \bar{u}(q\nu)\gamma_0\bm{\gamma}
        -
        \gamma_0\bm{\gamma}\upsilon(q_-\nu) \bar{\upsilon}(q_-\nu)\gamma_0\bm{\gamma}
    \bigg]
    \Bigg\}{u}(p\mu),
\end{multline}
\vspace{-5mm}
\begin{multline}\label{RVwithP}
    \frac{1}{2}\big[R^{(1)},V^{(1)}\big]_{b^\dagger b} = 
    \frac{m^2}{4(2\pi)^3}
    \int \frac{d \mathbf{p}}{E_{\mathbf{p}}^2}
     \frac{d \mathbf{q}}{E_{\mathbf{q}}\omega_{\mathbf{p}-\mathbf{q}}} b^\dagger(p\rho)b(p\rho) 
     \\ \times
     \bar{u}(p\mu)\Bigg\{
        \frac{P^{\alpha\beta}(k)}{E_{\mathbf{p}}-\omega_{\mathbf{p}-\mathbf{q}}-E_{\mathbf{q}}}\left[g\,\gamma_\alpha-\frac{f}{2m}ik^\xi\sigma_{\xi\alpha}\right] 
        u(q\nu) \bar{u}(q\nu) 
        \left[g\,\gamma_\beta-\frac{f}{2m}ik^\eta\sigma_{\beta\eta}\right]\\+
        \frac{P^{\alpha\beta}(k_-)}{E_{\mathbf{p}}+\omega_{\mathbf{p}-\mathbf{q}}+E_{\mathbf{q}}}\left[g\,\gamma_\alpha-\frac{f}{2m}ik^\xi_-\sigma_{\alpha\xi}\right] 
        \upsilon(q_-\nu) \bar{\upsilon}(q_-\nu) 
        \left[g\,\gamma_\beta-\frac{f}{2m}ik^\eta_-\sigma_{\eta\beta}\right]
    \Bigg\}{u}(p\mu).
\end{multline}
In these formulas $\cross{q}=q_\alpha\gamma^\alpha$, $q=(E_{\mathbf{q}},\mathbf{q})$, ${q_-=(E_{\mathbf{q}},-\mathbf{q})}$, and ${k \equiv (\omega_{\mathbf{p}-\mathbf{q}},\mathbf{p}-\mathbf{q})}$.
The r.h.s of Eq. (\ref{V2}) embodies merely the so-called contact terms in which instead of the propagators of intermediate particles we have the squares of their masses.
Keeping in mind the definition (\ref{Pk_nucl}) of the projection operator $P_{\alpha\beta}(k)$ we find
\begin{multline}\label{RV_nucl}
    \frac{1}{2}\big[R^{(1)},V^{(1)}\big]_{b^\dagger b}=
    -\frac{m^2}{4(2\pi)^3}
    \int \frac{d \mathbf{p}}{E_{\mathbf{p}}^2}
     \frac{d \mathbf{q}}{E_{\mathbf{q}}\omega_{\mathbf{p}-\mathbf{q}}}
     b^\dagger(p\rho)b(p\rho) 
     \\ \times
     \bar{u}(p\mu)\Bigg\{
        \frac{1}{E_{\mathbf{p}}-\omega_{\mathbf{p}-\mathbf{q}}-E_{\mathbf{q}}}\left[\gamma_\alpha(g+f)-\frac{f}{2m}(p+q)_\alpha\right] 
        u(q\nu) \bar{u}(q\nu)
        \left[\gamma^\alpha(g+f)-\frac{f}{2m}(p+q)^\alpha\right]\\+
        \frac{1}{E_{\mathbf{p}}+\omega_{\mathbf{p}-\mathbf{q}}+E_{\mathbf{q}}}\left[\gamma_\alpha(g+f)-\frac{f}{2m}(p-q_-)_\alpha\right] 
        \upsilon(q_-\nu) \bar{\upsilon}(q_-\nu)
        \left[\gamma^\alpha(g+f)-\frac{f}{2m}(p-q_-)^\alpha\right]
        \\+
        f \frac{E_{\mathbf{q}}}{m^2} [6m(g+f)+f(\mathbf{p}+\mathbf{q})\cdot\bm{\gamma}]
        +
        g^2 \frac{\omega_{\mathbf{p}-\mathbf{q}}}{m_b^2}
        \bigg[
            \gamma_0u(q\nu) \bar{u}(q\nu)\gamma_0
            -
            \gamma_0\upsilon(q_-\nu) \bar{\upsilon}(q_-\nu)\gamma_0
        \bigg]
        \\-
        f^2 \frac{\omega_{\mathbf{p}-\mathbf{q}}}{4m^2} 
        \bigg[
            \gamma_0\bm{\gamma}u(q\nu) \bar{u}(q\nu)\gamma_0\bm{\gamma}
            -
            \gamma_0\bm{\gamma}\upsilon(q_-\nu) \bar{\upsilon}(q_-\nu)\gamma_0\bm{\gamma}
        \bigg]
    \Bigg\}{u}(p\mu)
    \,.
\end{multline}
\end{widetext}
After substituting Eqs. (\ref{Mren_bb}), (\ref{V2}), and (\ref{RV_nucl}) to (\ref{ren_requirement}) we see that the last two contact terms in Eq. (\ref{RV_nucl}) cancels the contribution from Eq. (\ref{V2}). 

Doing so we get the nucleon mass shift due to the one-$\rho$-meson exchange
\begin{equation}\label{delta_mN_fin}
    \delta m^{(2)} \equiv \delta m_N^{(\textrm{1$\rho$-exc})}(p) = I_1(p)
    +I_2(p)\,,
\end{equation}
\begin{multline}\label{delta_mN_I1}
    I_1(p) = \frac{1}{m(2\pi)^3}\int \frac{d \mathbf{q}}{E_{\mathbf{q}}}\frac{1}{(p-q)^2-m_b^2}
    \bigg\{
        g^2(2m^2-pq)\\
        +3gf(m^2-pq)+\frac{f^2}{4m^2}(m^2-pq)(5m^2-pq)
    \bigg\}\,,
\end{multline}
\begin{multline}\label{delta_mN_I2}
    I_2(p) = \frac{1}{2m(2\pi)^3}\int \frac{d \mathbf{k}}{\omega_{\mathbf{k}}}\frac{1}{(p+k)^2-m^2}
    \bigg\{
        2g^2(m^2-pk)\\
        +3gf m_b^2+f^2\left(m_b^2+\frac{(pk)^2}{2m^2}\right)
    \bigg\}\,.
\end{multline}

In the case of QED one needs to put in Eqs. \eqref{ren_requirement}-\eqref{RVwithP} $g=e$, $f=0$, $m_b=\lambda$ (of course, now $m$ refers to the electron mass) and 
\begin{multline}\label{VCoul}
    V^{(2)}_{b^\dagger b} 
    \equiv V_\textrm{Coul}\,_{b^\dagger b} = 
    \frac{e^2 m^2}{4(2\pi)^3}
    \int \frac{d \mathbf{p}}{E_{\mathbf{p}}^2}
    \frac{d \mathbf{q}}{E_{\mathbf{q}}\omega_{\mathbf{p}-\mathbf{q}}}
    \\ \times
    b^\dagger(p\rho)b(p\rho)
    \frac{\omega_{\mathbf{p}-\mathbf{q}}}{(\mathbf{p}-\mathbf{q})^2+\lambda^2}
    \bar{u}(p\mu)
    \\ \times
    \bigg[ 
        \gamma_0u(q\nu) \bar{u}(q\nu)\gamma_0
        -
        \gamma_0\upsilon(q_-\nu) \bar{\upsilon}(q_-\nu)\gamma_0
    \bigg]
    {u}(p\mu).
\end{multline}
The polarization vectors $e(k\sigma)$ ($\sigma=\pm 1$) in (\ref{varphi_expan}) have the properties (in the Coulomb gauge) $\mathbf{k}\cdot\mathbf{e}(k\sigma)=0$, ${e_0(k\sigma)=0}$, being normalized as \cite{Wein1}
\begin{multline}\label{Pk}
    \sum_\sigma e_\alpha(k\sigma)e_\beta^*(k\sigma) =
    -g_{\alpha\beta} 
    +k_0\frac{k_\alpha n_\beta + k_\beta n_\alpha}{\mathbf{k}^2} 
    \\
    -\frac{k_\alpha k_\beta}{\mathbf{k}^2}
    -\frac{k^2}{\mathbf{k}^2}n_\alpha n_\beta
    \equiv P_{\alpha\beta}(k),
\end{multline}
with the timelike vector $ n=(1,0,0,0) $.
By using this definition we come to
\begin{multline}\label{RV_qed}
    \frac{1}{2}\big[R^{(1)},V^{(1)}\big]_{b^\dagger b}=
    -\frac{e^2 m^2}{4(2\pi)^3}
    \int \frac{d \mathbf{p}}{E_{\mathbf{p}}^2}
    \frac{d \mathbf{q}}{E_{\mathbf{q}}\omega_{\mathbf{p}-\mathbf{q}}}
    \\ \times
    b^\dagger(p\rho)b(p\rho)
     \bar{u}(p\mu)\Bigg\{
        \frac{1}{E_{\mathbf{p}}-\omega_{\mathbf{p}-\mathbf{q}}-E_{\mathbf{q}}} \gamma_\alpha  
        u(q\nu) \bar{u}(q\nu)
        \gamma^\alpha
        \\+
        \frac{1}{E_{\mathbf{p}}+\omega_{\mathbf{p}-\mathbf{q}}+E_{\mathbf{q}}} \gamma_\alpha  
        \upsilon(q_-\nu) \bar{\upsilon}(q_-\nu)
        \gamma^\alpha
        +
        \frac{\omega_{\mathbf{p}-\mathbf{q}}}{(\mathbf{p}-\mathbf{q})^2}
        \\\times
        \bigg[
            \gamma_0u(q\nu) \bar{u}(q\nu)\gamma_0
            -
            \gamma_0\upsilon(q_-\nu) \bar{\upsilon}(q_-\nu)\gamma_0
        \bigg]
    \Bigg\}{u}(p\mu).
\end{multline}
The important point is that the contact terms cancel again (cf. the last term in Eq. \eqref{RV_qed} with contribution from Eq. \eqref{VCoul}).
After it the electron mass shift due to the one-photon exchange can be written as
\begin{multline}\label{delta_me-_for_diagrams}
    \delta m^{(2)} \equiv \delta m_{e^-}^{(\textrm{1ph-exc})}(p) 
    =
    \frac{e^2}{8(2\pi)^3}
    \int \frac{d \mathbf{q}}{E_{\mathbf{q}}\omega_{\mathbf{p}-\mathbf{q}}}\bar{u}(p\mu)\gamma_\alpha
    \\ \times
    \bigg\{
    \frac{\cross{q} + m}{E_{\mathbf{p}}-\omega_{\mathbf{p}-\mathbf{q}}-E_{\mathbf{q}}}+
    \frac{\cross{q}_- - m}{E_{\mathbf{p}}+\omega_{\mathbf{p}-\mathbf{q}}+E_{\mathbf{q}}}
    \bigg\}
    \gamma^\alpha{u}(p\mu)
    \\=
    \frac{e^2}{m (2\pi)^3}
    \int \frac{d \mathbf{q}}{2 E_{\mathbf{q}} \omega_{\mathbf{p}-\mathbf{q}}}
    \bigg\{
        \frac{2m^2-pq}{E_{\mathbf{p}}-\omega_{\mathbf{p}-\mathbf{q}}-E_{\mathbf{q}}}
        \\-
        \frac{2m^2+pq_-}{E_{\mathbf{p}}+\omega_{\mathbf{p}-\mathbf{q}}+E_{\mathbf{q}}}
    \bigg\}.
\end{multline}
The last expression can be represented in an explicitly covariant form by using the trick from \cite{KorShe2004} (see Appendix therein), with help of the relation
\begin{multline}
    \int\frac{d \mathbf{q}}{2E_{\mathbf{q}}\omega_{\mathbf{p}-\mathbf{q}}}
    \bigg\{
        \frac{A(pq)}{E_{\mathbf{p}}-\omega_{\mathbf{p}-\mathbf{q}}-E_{\mathbf{q}}}
        -
        \frac{B(pq_-)}{E_{\mathbf{p}}+\omega_{\mathbf{p}-\mathbf{q}}+E_{\mathbf{q}}}
    \bigg\}
    \\=
    \int \frac{d \mathbf{q}}{E_{\mathbf{q}}}\frac{A(pq)}{(p-q)^2-m_b^2}
    +
    \int \frac{d \mathbf{k}}{\omega_{\mathbf{k}}}\frac{C(pk)}{(p+k)^2-m^2}
    \\=
    \int \frac{d \mathbf{q}}{E_{\mathbf{q}}}\frac{B(pq)}{(p+q)^2-m_b^2}
    +
    \int \frac{d \mathbf{k}}{\omega_{\mathbf{k}}}\frac{D(pk)}{(p-k)^2-m^2}\,,
\end{multline}
where covariant numerators $A(pq)$, $B(pq)$, $C(pk)$, and $D(pk)$ should be connected by
\begin{align}
    A(pq_-)(E_{\mathbf{p}}+\omega_{\mathbf{p}+\mathbf{q}}+E_{\mathbf{q}}) &- 
    B(pq)(E_{\mathbf{p}}+\omega_{\mathbf{p}+\mathbf{q}}-E_{\mathbf{q}})
    \nonumber
    \\
    &\hspace{15mm} = 2E_{\mathbf{q}} C(pk')\,,
    \nonumber
    \\
    A(pq)(E_{\mathbf{p}}-\omega_{\mathbf{p}-\mathbf{q}}+E_{\mathbf{q}}) &- 
    B(pq_-)(E_{\mathbf{p}}-\omega_{\mathbf{p}-\mathbf{q}}-E_{\mathbf{q}})
    \nonumber
    \\
    &\hspace{15mm} = 2E_{\mathbf{q}} D(pk)\,,
\end{align}
with ${k'=(\omega_{\mathbf{p}+\mathbf{q}},-\mathbf{p}-\mathbf{q})}$ and ${k=(\omega_{\mathbf{p}-\mathbf{q}},\mathbf{p}-\mathbf{q})}$.
Doing so, we get
\begin{equation}\label{deltamUCT}
    \delta m_{e^-}^{(\textrm{1ph-exc})}(p) = I'_1(p)+I'_2(p)\,,
\end{equation}
\begin{equation}\label{delta_me-_I1}
    I'_1(p) = \frac{e^2}{m(2\pi)^3}\int \frac{d \mathbf{q}}{E_{\mathbf{q}}}\frac{2m^2-pq}{(p-q)^2-\lambda^2},
\end{equation}
\begin{equation}\label{delta_me-_I2}
    I'_2(p) = \frac{e^2}{m(2\pi)^3}\int \frac{d \mathbf{k}}{\omega_{\mathbf{k}}}\frac{m^2-pk}{(p+k)^2-m^2}
\end{equation}
or in the symmetrized form
\begin{multline}\label{deltamUCT_symmetrized}
    \delta m_{e^-}^{(\textrm{1ph-exc})}(p) =
    \frac{e^2}{2m(2\pi)^3}
    \\\times
    \bigg\{
        \int \frac{d \mathbf{q}}{E_{\mathbf{q}}}
        \bigg[
        \frac{2m^2-pq}{(p-q)^2-\lambda^2}
        +
        \frac{2m^2+pq}{(p+q)^2-\lambda^2}
        \bigg]
        \\+
        \int \frac{d \mathbf{k}}{\omega_{\mathbf{k}}}
        \bigg[
        \frac{m^2+pk}{(p-k)^2-m^2}
        +
        \frac{m^2-pk}{(p+k)^2-m^2}
        \bigg]
    \bigg\}.
\end{multline}
A similar trick taken from our previous paper \cite{KorShe2004} (see Appendix therein) has been used to get the covariant integrals \eqref{delta_mN_I1} and \eqref{delta_mN_I2}, i.e., $I_{1,2}(p)=I_{1,2}(\Lambda p)$.
The latter simplifies the further calculations $I_{1,2}(p)=I_{1,2}(m,0,0,0) ~ \forall p$, reducing integrals \eqref{delta_me-_I1}, \eqref{delta_me-_I2} to quadratures:
\begin{multline}
    I'_{1}(p)=I'_{1}(m,0,0,0)
    \\=
    m\frac{e^2}{(2\pi)^2}
    \lim_{\lambda \rightarrow 0}\int\limits_0^\infty d t
    \frac{t^2}{\sqrt{t^2+1}}\frac{2-\sqrt{t^2+1}}{1-\frac{1}{2}\lambda -\sqrt{t^2+1}},
\end{multline}
\begin{multline}
    I'_{2}(p)=I'_{2}(m,0,0,0)
    \\=
    m\frac{e^2}{(2\pi)^2}
    \lim_{\lambda \rightarrow 0}\int\limits_0^\infty d t
    \frac{t^2}{\sqrt{t^2+\lambda }}\frac{1-\sqrt{t^2+\lambda }}{\frac{1}{2}\lambda +\sqrt{t^2+\lambda }}.
\end{multline}
Trying to cure the drawback of local field theories, these divergent integrals can be regularized by introducing the corresponding cutoff functions. We will consider such a possibility in the next section.

By the way, it is easy to see that the integrals (\ref{delta_mN_I1}) and (\ref{delta_mN_I2}) become equal, respectively, to (\ref{delta_me-_I1}) and (\ref{delta_me-_I2}) if we put $g=e$, $f=0$, and $m_b=\lambda$.

\begin{figure}[t]
    \includegraphics[scale=0.4]{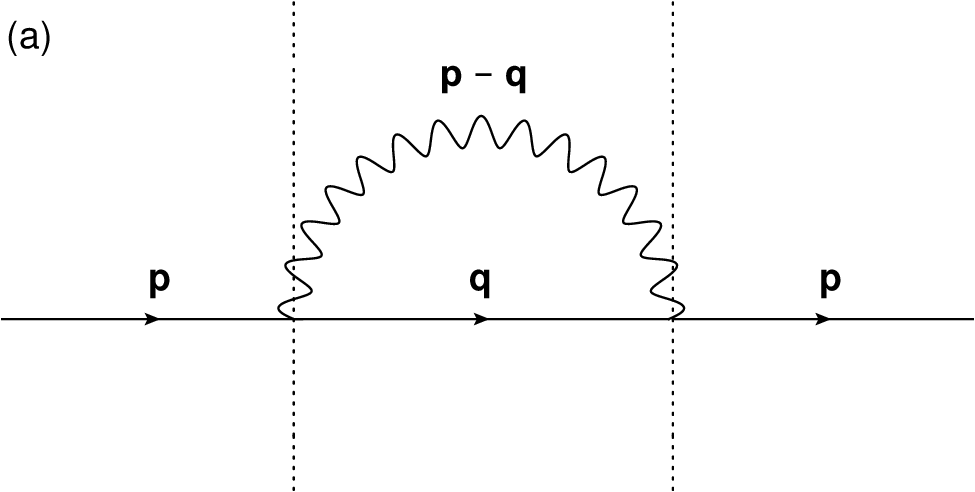}
    \includegraphics[scale=0.4]{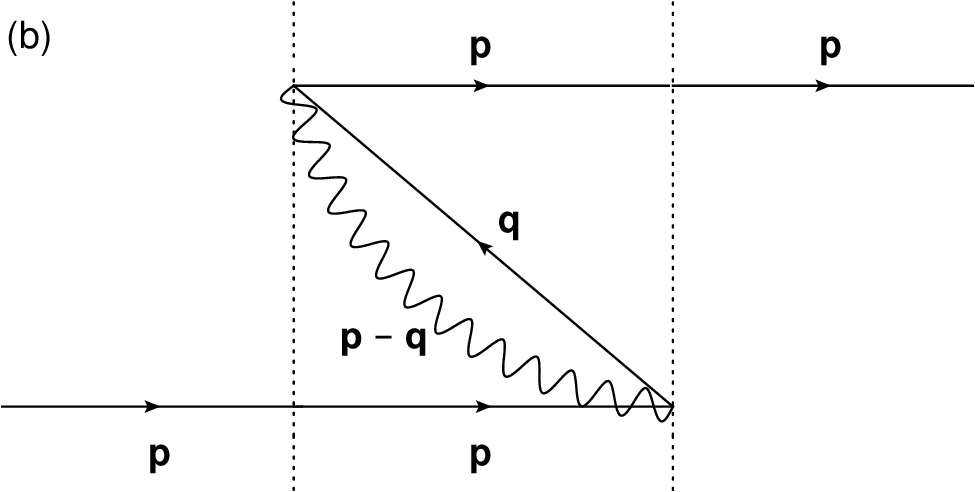}
    \caption{Two contributions to the mass shift within the Old-Fashioned Perturbation
    Theory \label{Old_fashioned_diagrams}}
\end{figure}

Separate contributions in the curly brackets of (\ref{delta_me-_for_diagrams}) can be represented via the graphs (a) and (b) in Fig. \ref{Old_fashioned_diagrams}. Such graphs are typical of the old-fashioned perturbation theory. In this context, the inverse energy denominators in the r.h.s. of Eq. (\ref{delta_me-_for_diagrams}) have the form $D^{-1}(E)=(E-E_i)^{-1}$ with the appropriate values of the collision energy $E$ and energy of all permissible intermediate states $E_i$. In other words, $ (E_{\mathbf{p}}-\omega_{\mathbf{p}-\mathbf{q}}-E_{\mathbf{q}})^{-1} $ and $ (E_{\mathbf{p}}+\omega_{\mathbf{p}-\mathbf{q}}+E_{\mathbf{q}})^{-1} $ are related to the propagators
\begin{equation}
    \begin{split}
        &D^{-1}(E=E_{\mathbf{p}})=(E-\omega_{\mathbf{p}-\mathbf{q}}-E_{\mathbf{q}})^{-1} ~ \textrm{and}
        \\
        &D^{-1}(E=E_{\mathbf{p}})=(E-E_{\mathbf{p}}-\omega_{\mathbf{p}-\mathbf{q}}-E_{\mathbf{q}}-E_{\mathbf{p}})^{-1}\,,
    \end{split}
\end{equation}
being associated with the two and four internal lines between the dotted (the phantoms if one uses the terminology adopted in Ref. \cite{KoCaSh2007}), respectively, in graphs $a$ and $b$.

The graphs in Fig. \ref{Old_fashioned_diagrams} are topologically equivalent to
the time-ordered Feynman diagram Fig. \ref{Feyn_self_energy_diagram}. However, in the Schr\"odinger picture employed here, where all events are related to one and the same instant, such an analogy seems to be misleading. In fact, the line directions in Fig. \ref{Old_fashioned_diagrams} are given with the sole scope to discriminate between the fermion and antifermion states.

At the end of this section, we would like to present the result obtained previously \cite{KorShe2004} for the mass shift of the nucleon due to the one-pion exchange
\begin{multline}
    \delta m_{N}^{(\textrm{1$\pi$-exc})}(p)
    =
    \frac{g_\pi^2}{2m(2\pi)^3}
    \bigg\{
        \int \frac{d \mathbf{q}}{E_{\mathbf{q}}}
        \frac{m^2-pq}{(p-q)^2-m_\pi^2}
        \\+
        \int \frac{d \mathbf{k}}{\omega_{\mathbf{k}}}
        \frac{-pk}{(p+k)^2-m^2}
    \bigg\}
    \\=
    \frac{g_\pi^2}{4m(2\pi)^3}
    \bigg\{
        \int \frac{d \mathbf{q}}{E_{\mathbf{q}}}
        \bigg[ 
            \frac{m^2-pq}{(p-q)^2-m_\pi^2} +\frac{m^2+pq}{(p+q)^2-m_\pi^2}
        \bigg]
        \\+
        \int \frac{d \mathbf{k}}{\omega_{\mathbf{k}}}
        \bigg[ 
            \frac{pk}{(p-k)^2-m^2} 
            +\frac{-pk}{(p+k)^2-m^2}
        \bigg]
    \bigg\}.
\end{multline}
We see that it has the same denominators as in Eqs. \eqref{delta_mN_fin}-\eqref{delta_mN_I2} and \eqref{deltamUCT_symmetrized} but the numerators are different.
If one considers model with interacting nucleons, pions, $\rho$ mesons, and other mesons, then the nucleon mass shift would be a sum $\delta m_N = \delta m_{N}^{({1\pi\textrm{-exc}})} + \delta m_{N}^{({1\rho\textrm{-exc}})} + \dots$.

\section{Regularization of the field models with Yukawa-type couplings vs the problem of ultraviolet divergences\label{Cutoff}}

Evidently, because of insufficiently fast falloff of the integrands in Eqs. \eqref{delta_mN_I1}-\eqref{delta_mN_I2} and \eqref{delta_me-_I1}-\eqref{delta_me-_I2}, when the momentum absolute value increasing, one has to handle ultraviolet divergent quantities.
Thus, it is necessary to regularize the integrals, so they take on finite values. 
Trying to overcome the divergences we would like to propose a nonlocal extension of the field models in question. 

Let us come back to the division
\begin{equation}
    H(\alpha)=K(\alpha_c)=H_F(\alpha_c)+K_I(\alpha_c).
\end{equation}
It is the place where we would like to stress that the interaction $K_I(\alpha_c)$ between the clothed particles does not contain the contact terms. This suggestion is valid, at least, in the second order in the coupling constants. In other words, the interactions in the r.h.s. of \eqref{KI} are free from such terms. 
To verify this for the case of QED, one can refer to work \cite{ArsKosShe2021}, where we have presented the derivation of all two-particle interactions $K(e^-e^- \rightarrow e^-e^-)$, $K(e^+e^+ \rightarrow e^+e^+)$, $K(e^-e^+ \rightarrow e^-e^+)$, $K(\gamma e^- \rightarrow \gamma e^-)$, $K(\gamma e^+ \rightarrow \gamma e^+)$, $K(\gamma\gamma \rightarrow e^-e^+)$, $K(e^-e^+ \rightarrow \gamma\gamma)$, accessible in our case.
The derivation of $K(NN \rightarrow NN)$ and $K(N\bar{N} \rightarrow N\bar{N})$ operators in the theory with nucleon-$\rho$-meson coupling is presented in the Appendix \ref{KNN_derivation} (see also \cite{SheDub2010}).

In this context, we recall the transformation properties of the $S$-matrix 
$\langle f |S| i  \rangle$ 
in the D-picture
with respect to the Lorentz group
\begin{equation}\label{Smatrix_property}
    \langle f |S| i  \rangle =
    \langle \Lambda f |S| \Lambda i  \rangle, ~ \forall \Lambda
\end{equation}
(cf. the discussion on the p. 83 in Ref. \cite{GoldWat} and, in particular, Eq. (103) therein).
It is well known that such property is provided if the interaction Hamiltonian density $H_I(x)$ in the D-picture meets 
\begin{equation}\label{HI_scalar}
    U_F(\Lambda)H_{I}(x)U_F^{-1}(\Lambda) = H_{I}(\Lambda x).
\end{equation}
In the theories with interacting fermions and scalar bosons the property \eqref{HI_scalar} of the $H_I$ density to be the Lorentz scalar fulfills even in the BPR. But in the models with vector bosons the operator $H_I(x)$ embodies the terms ($V^{(2)}(x)$ or $V_{\textrm{Coul}}(x)$) that are not Lorentz scalars.
These noncovariant contributions no longer present in the Hamiltonian. It is a distinctive feature of the CPR.

Therefore, we choose the interaction $K_I^{(2)}$ as a starting point for our nonlocal extension of the model. We introduce the cutoff factors which are supposed to account for finite-size effects. It can be achieved if in the r.h.s. of \eqref{KI} we replace the structures
\begin{equation}
    \begin{split}
        &\bar{u}(p_1\mu_1)\Gamma u(p_2\mu_2),~
        \bar{u}(p_1\mu_1)\Gamma \upsilon(p_2\mu_2),\\
        &\bar{\upsilon}(p_1\mu_1)\Gamma u(p_2\mu_2),~
        \bar{\upsilon}(p_1\mu_1)\Gamma \upsilon(p_2\mu_2)
    \end{split}
\end{equation}
($\Gamma$ is some combination of the gamma matrices) with ones multiplied by the cutoff factors $g_{11}(p_1,p_2)$, $g_{12}(p_1,p_2)$, $g_{21}(p_1,p_2)$, and $g_{22}(p_1,p_2)$, respectively. 
Along the guideline we replace the interaction Hamiltonian in the CPR \eqref{KI} by
\begin{multline}\label{KI_nloc}
    K^{\textrm{nloc}\,(2)}_{I}=K^\textrm{nloc}(ff \rightarrow ff) + K^\textrm{nloc}(\bar{f}\bar{f} \rightarrow \bar{f}\bar{f}) 
    \\+
    K^\textrm{nloc}(f\bar{f} \rightarrow f\bar{f}) 
    + K^\textrm{nloc}(bf \rightarrow bf) + \cdots
    \\
    + K^{\textrm{nloc}\,(2)}_\textrm{1-body}
    + M^{\textrm{nloc}\,(2)}_\textrm{ren}
    .
\end{multline}

In order to fulfill the property \eqref{Smatrix_property} it is required (see details in the Appendix \ref{KNN_derivation})
\begin{equation}
    g_{\varepsilon' \varepsilon}(\Lambda p_1, \Lambda p_2)=g_{\varepsilon' \varepsilon}(p_1, p_2) ~~~ (\varepsilon',\varepsilon = 1,2),
\end{equation}
i.e., these cutoff factors should be dependent on the Lorentz scalar $p_1p_2$.

By assumption, our interactions are invariant with respect to the space inversion $P$, charge conjugation $C$, and time reversal $\mathcal{T}$. If the modified interaction retain such invariance, then the following relations take place
\begin{equation}
    \begin{split}
        &g_{11}(p_1,p_2)=g_{22}(p_2,p_1),\\
        &g_{12}(p_1,p_2)=g_{21}(p_2,p_1),\\
        &g_{\varepsilon' \varepsilon}(p_1,p_2)=g_{\varepsilon' \varepsilon}(p_{1-},p_{2-}).
    \end{split}
\end{equation}

Drawing parallels with the mass counterterm \eqref{Mfermion_def} in the local field model we consider its nonlocal extension
\begin{equation}
    M^{\textrm{nloc}\,(2)}_\textrm{ferm}
    =m \int \frac{d\mathbf{p}}{E_\mathbf{p}^2} 
    F^\dagger(p\mu_1) M^\textrm{nloc}(p\mu_1\mu_2) F(p\mu_2),
\end{equation}
where one has to handle the matrix 
\begin{multline*}
    M^\textrm{nloc}(p\mu_1\mu_2)\\= 
    \begin{bmatrix}
        m^{(2)}_{11}(p)\delta_{\mu_1\mu_2} & 
        m^{(2)}_{12}(p)\bar u(p\mu_1) \upsilon(p_-\mu_2)\\ 
        m^{(2)}_{21}(p)\bar \upsilon(p_-\mu_1) u(p\mu_2) & 
        -m^{(2)}_{22}(p)\delta_{\mu_1\mu_2}
    \end{bmatrix}.
\end{multline*}
Here the coefficients $m_{\varepsilon' \varepsilon}(p)$ may be momentum dependent. Of course, for simplicity, real $m_{12}(p)$ and $m_{21}(p)$ should be equal, to ensure the hermiticity of the operator $M^{\textrm{nloc}\,(2)}_\textrm{ferm}$.

It is proved that a considerable part of the calculations, presented in Sec. \ref{mass_shifts_UCT}, remains intact with the modified interactions.
In particular, the corresponding regularization of the operator $K^{(2)}_\textrm{1-body}\,_{b^\dagger b} \equiv V^{(2)}_{b^\dagger b} + \frac{1}{2}\big[R^{(1)}, V^{(1)}\big]_{b^\dagger b}$ (defined by Eqs. \eqref{VCoul} and \eqref{RV_qed}) gives
\begin{multline}
    K^{\textrm{nloc}\,(2)}_\textrm{1-body}\,_{b^\dagger b} =
    -\frac{e^2 m^2}{4(2\pi)^3}
    \int \frac{d \mathbf{p}}{E_{\mathbf{p}}^2}
    \frac{d \mathbf{q}}{E_{\mathbf{q}}\omega_{\mathbf{p}-\mathbf{q}}}
    b^\dagger(p\rho)b(p\rho)
    \\ \times
    \bar{u}(p\mu)\Bigg\{
        \frac{g_{11}^2(pq)}{E_{\mathbf{p}}-\omega_{\mathbf{p}-\mathbf{q}}-E_{\mathbf{q}}} \gamma_\alpha  
        u(q\nu) \bar{u}(q\nu)
        \gamma^\alpha
        \\+
        \frac{g_{21}^2(pq_-)}{E_{\mathbf{p}}+\omega_{\mathbf{p}-\mathbf{q}}+E_{\mathbf{q}}} \gamma_\alpha  
        \upsilon(q_-\nu) \bar{\upsilon}(q_-\nu)
        \gamma^\alpha
        \\+
        \frac{\omega_{\mathbf{p}-\mathbf{q}}}{(\mathbf{p}-\mathbf{q})^2}
        \bigg[
            g_{11}^2(pq) \gamma_0u(q\nu) \bar{u}(q\nu)\gamma_0
            \\-
            g_{21}^2(pq_-) \gamma_0\upsilon(q_-\nu) \bar{\upsilon}(q_-\nu)\gamma_0
        \bigg]
    \Bigg\}{u}(p\mu).
\end{multline}
In agreement with the first requirement to the clothing procedure proposed in (p. 6 on Ref. \cite{SheShi2001}) the latter should be compensated by choosing the coefficient $m^{(2)}_{11}(p)$ such that 
$M^{(2)}_\textrm{ferm}\,_{b^\dagger b} = - K^{\textrm{nloc}\,(2)}_\textrm{1-body}\,_{b^\dagger b}$,
so we arrive to a nonlocal analogue of the electron mass shift\footnote{Of course, in general, each coefficient $m^{(2)}_{\varepsilon' \varepsilon}(p)$ is defined 
with an accuracy to adding a function $f_{\varepsilon' \varepsilon}(p)$ such that $\int f_{\varepsilon' \varepsilon}(p) d\mathbf{p}/E_{\mathbf{p}}^2=0$.}
\begin{multline}\label{nonloc_deltam}
    m^{(2)}_{11}(p)
    =
    \frac{e^2}{m (2\pi)^3}
    \int \frac{d \mathbf{q}}{2 E_{\mathbf{q}} \omega_{\mathbf{p}-\mathbf{q}}}
    \\\times
    \bigg\{
        g^2_{11}(pq)\frac{2m^2-pq}{E_{\mathbf{p}}-\omega_{\mathbf{p}-\mathbf{q}}-E_{\mathbf{q}}}
        \\-
        g^2_{21}(pq_-)\frac{2m^2+pq_-}{E_{\mathbf{p}}+\omega_{\mathbf{p}-\mathbf{q}}+E_{\mathbf{q}}}
    \bigg\}.
\end{multline}
Since the structure of the integral (76) is similar to the r.h.s. of Eq. (38), we can repeat those transformations to formula (44), in order to find the expression with Feynman-like propagators
\begin{multline}
    m^{(2)}_{11}(p)=
    \frac{e^2}{2m(2\pi)^3}
    \bigg\{
        \int \frac{d \mathbf{q}}{E_{\mathbf{q}}}
        \bigg[
            g_{11}^2(pq) \frac{2m^2-pq}{(p-q)^2-\lambda^2}
            \\+
            g_{21}^2(pq) \frac{2m^2+pq}{(p+q)^2-\lambda^2}
        \bigg]
        +
        \int \frac{d \mathbf{k}}{\omega_{\mathbf{k}}}
        \bigg[
            g_{11}^2(pq') \frac{m^2+pk}{(p-k)^2-m^2}
            \\+
            g_{21}^2(pq'_-)\frac{m^2-pk_-}{(p+k_-)^2-m^2}
        \bigg]
        +
        \int \frac{d \mathbf{q}}{E_{\mathbf{q}}}
        \big(g_{21}^2(pq) - g_{11}^2(pq_-)\big)
        \\\times
        \frac{E_{\mathbf{p}}}{\omega_{\mathbf{p}+\mathbf{q}}}
        \frac{m^2+qk'}{(q-k')^2-m^2}
    \bigg\}.
\end{multline}
Here ${q'=(E_{\mathbf{p}-\mathbf{k}},\mathbf{p}-\mathbf{k})}$ and ${k'=(\omega_{\mathbf{p}+\mathbf{q}},\mathbf{p}+\mathbf{q})}$.
Obviously, if one puts $g^2_{11}=g^2_{21}=1$ we come back to the local result \eqref{deltamUCT_symmetrized}.

Unlike the momentum-independent mass shift \eqref{deltamUCT} obtained in the local model this coefficient may be depended on the momentum. But the most significant property of the integrals \eqref{nonloc_deltam} is to take on finite values. This property leads to the total cancellation of the finite terms $K^{(2)}_\textrm{1-body}\,_{b^\dagger b} $ and $M^{(2)}_\textrm{ren}\,_{b^\dagger b}$ in the Hamiltonian \eqref{KI_nloc}.
At the point one should realize that within our approach, the one-particle operators cannot appear in the new form $K(\alpha_c)$ of the Hamiltonian.

One should stress that the integral \eqref{nonloc_deltam} with properly selected cutoffs $g_{11}$ , $g_{21}$ takes on finite values for non-zero $\lambda$ in $\omega_{\mathbf{p}-\mathbf{q}}=\sqrt{\lambda^2+(\mathbf{p}-\mathbf{q})^2}$.
In other words, the two last terms in the decomposition \eqref{KI_nloc} are cancelled. Thus, the pleasant feature of the CPR with $\lambda \ne 0$ takes place as before.
Otherwise, we will encounter the infrared singularity. 

Indeed, the parameter $\lambda$ introduced in Eq. \eqref{VCoul_def} has to be put zero at the end of calculations, viz., when evaluating some observables or matrix elements $\langle f|S|i \rangle$. But in the theory developed here quantities like $m_{11}^{(2)}(p)$ or $\delta m^{(2)}$ are not considered as some corrections to the badly defined bare mass $m_0$. Instead, they are introduced as free parameters that should be chosen to cancel the terms in $H$ that prevent Eq. \eqref{mass_def} to be valid.
So, the cancellation of the one-body terms happens in the Hamiltonian itself. Therefore, we do not encounter infrared singularities considering the problem of fermion mass renormalization.

There are no counterterms such as $M^{(2)}_\textrm{ren}$ to cancel singularities present in the two-body terms of the Hamiltonian~\eqref{KI_nloc}. This issue will be addressed our upcoming research where such singularities arise when evaluating the positronium wave functions using the electron-positron interaction operator $K(e^- e^+ \to e^- e^+)$. Fortunately, Lande's technique \cite{KwoTab1978} can be used to overcome these singularities when solving the eigenvalue equations, e.g., for the Coulomb problem in momentum space.
The realization of the Lande's idea is underway.

In Ref. \cite{FroShe2012} it is shown that the removal of the mass counterterms is closely interwoven with ensuring the relativistic invariance of the theory as a whole. In this connection, one should note that the supplementary conditions on the choice of $g$-cutoffs are imposed by the commutations of the Hamiltonian with the boost generators $[\mathbf{B},H]=i\mathbf{P}$. We are not going to construct some special forms of these cutoffs here. In many applications the parameters involved in $g_{\varepsilon' \varepsilon}(p_1,p_2)$ are chosen in order to ensure the best fit of the available data, see the example in Ref.~\cite{SheDub2010}.

\section{ Comparison with Dyson-Feynman approach: elimination of divergences in the $S$-matrix\label{comparison}}

We have seen that the procedure developed enables us to remove from the Hamiltonian in the CPR not only the ``bad'' terms (at least, up to any given order in the coupling constants). Simultaneously, the "good" one-body terms are eliminated too being compensated with the corresponding mass counterterms. By the way, it means that the corresponding divergences in the conventional form of $H$ can no longer appear in the $S$-matrix. 
In this section we will reproduce our results \eqref{delta_mN_fin} and \eqref{deltamUCT}, relying upon the Dyson-Feynman series for the $S$ operator, viz.,
\begin{multline}\label{Sexp}
    S(\alpha)
    =1 + S^{I} + S^{II} + \cdots
    =\sum_{n=0}^\infty \frac{(-i)^n}{n!}
    \\\times
    \int\limits^\infty_{-\infty}dt_1\cdots\int\limits^\infty_{-\infty}dt_nP[H_I(t_1)\cdots H_I(t_n)],
\end{multline}
where $H_I(t)=e^{iH_F t} H_I e^{-iH_F t}$ is an interaction in the D-picture. Unlike the superscripts in $S^{(n)}$ the Roman indices in $S^{I}$, $S^{II}$, etc., denote the order of their appearance in the series \eqref{Sexp}.
The operator $H_I$ stem from the decomposition of the total Hamiltonian $H(\alpha)=H_F(\alpha)+H_I(\alpha)$ and is determined by Eq. \eqref{HI_def}. 
To be definite, we consider the matrix elements $\langle f|S^{(2)}|i \rangle$ of the second-order in the coupling constants $S$ operator between the initial and final boson-fermion states.
Following \cite{KoCaSh2003,She2004} one can derive 
\begin{equation*}
    \langle a_c^\dagger \cdots \Omega|S(\alpha_c)|a_c^\dagger \cdots \Omega \rangle =
    \langle a^\dagger \cdots \Omega_0|S(\alpha)|a^\dagger \cdots \Omega_0 \rangle
\end{equation*}
since the $\alpha$-algebra is isomorphic to the $\alpha_c$-algebra.
It allows us to handle the operator $S(\alpha_c)$ that is expressed in terms of a set $\{\alpha_c\}$, unlike the operator $S(\alpha)$ in the BPR, along with the clothed states
\begin{equation}\label{f_and_i_states}
    \begin{split}
        &|i \rangle = a_c^\dagger (k\sigma) b_c^\dagger (p\mu) |\Omega \rangle,
        \\
        &|f \rangle = a_c^\dagger (k'\sigma') b_c^\dagger (p'\mu') |\Omega \rangle.
    \end{split}
\end{equation}

According to the equivalence theorem \cite{She2004}, we have the relation $S_\textrm{cloth} = S(\alpha_c)$
since the UCT operator in the D-picture
\begin{equation}
    W_D(t) = e^{iK_F t} W e^{-iK_F t}
\end{equation}
meets the condition
\begin{equation}\label{equiv_condition}
    \lim_{t \to \pm \infty} W_D(t) = 1.
\end{equation}
So we have the equality
\begin{equation*}
    \langle a_c^\dagger \cdots \Omega|S_\textrm{cloth}|a_c^\dagger \cdots \Omega \rangle =
    \langle a^\dagger \cdots \Omega_0|S(\alpha)|a^\dagger \cdots \Omega_0 \rangle.
\end{equation*}
Here the operator $S_\textrm{cloth}$ corresponds to the division of the Hamiltonian $H=K_F(\alpha_c)+K_I(\alpha_c)$.

Being taken between the states \eqref{f_and_i_states} only terms of the type $b^\dagger b$, $a^\dagger a$, and $a^\dagger b^\dagger a b$ contribute to the matrix elements
\begin{multline}
    \langle f| S^{(2)} (\alpha_c) |i \rangle=
    \langle f| \bigg\{
        -i\int^\infty_{-\infty}\, dt\,e^{iK_F t} \big( M^{(2)}_\textrm{ren} 
        \\
        + V^{(2)} \big) e^{-iK_F t}
        + S_\textrm{SE}^{(2)} + S^{(2)}(bf \rightarrow bf)
    \bigg\} |i \rangle,
\end{multline}
where
\begin{multline}\label{Sse_def}
    S_\textrm{SE}^{(2)}= -\frac{1}{2}
    \int\limits^\infty_{-\infty}dt_1\int\limits^\infty_{-\infty}dt_2\,P[V^{(1)}(t_1)V^{(1)}(t_2)]_\textrm{one-body}
    \\
    =S_\textrm{fSE}^{(2)} +S_\textrm{bSE}^{(2)}
\end{multline}
determines the so-called self-energy operator
and $S^{(2)}(bf \rightarrow bf)$ arises from the third term ($S^{II}$) of the Dyson-Feynman series.

\begin{figure}
    \includegraphics[scale=0.4]{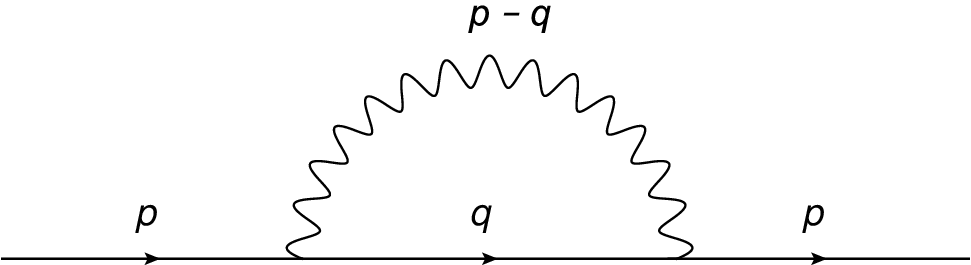}
    \caption{Self-energy Feynman diagram \label{Feyn_self_energy_diagram}}
\end{figure}

At this point we will consider the interplay between the $b^\dagger b$ type components of the operator $S^{(2)}$.
The ``forward-scattering'' process associated with the latter would be responsible for the appearance of certain infinity in the boson-fermion scattering amplitude $\langle f| \tau |i \rangle$.
Following a common practice, these terms should be compensated
\begin{multline}\label{ren_requirement_FD}
    S^{(2)}_{b^\dagger b}=
    -i\int^\infty_{-\infty}\, dt\,e^{iK_F t} \big( M^{(2)}_\textrm{ren}\,_{b^\dagger b} + V^{(2)}\,_{b^\dagger b} \big) e^{-iK_F t}\\
    + S_\textrm{fSE}^{(2)} =0
\end{multline}
so
\begin{multline}\label{ren_old_requirement_FD}
    2\pi i \delta(E_f-E_i)\langle f|( M^{(2)}_\textrm{ren}\,_{b^\dagger b} 
    \\
    + V^{(2)}\,_{b^\dagger b} )|i \rangle =
    \langle f| S_\textrm{fSE}^{(2)} |i \rangle.
\end{multline}
Note that the same equation would be for antifermion $d^\dagger d$ terms if one considered the matrix elements between boson-antifermion states. 
The fermion self-energy operator can be written in the form
\begin{equation}\label{Sse_Use}
    S_\textrm{fSE}^{(2)} \equiv -i\int\limits^\infty_{-\infty}\, dt\,e^{iK_F t} U_\textrm{fSE}^{(2)} e^{-iK_F t}.
\end{equation}
The property 
\begin{equation}\label{eA_B_eA}
    e^{-iK_F t} b(p\mu) e^{iK_F t}=b(p\mu) e^{iE_{\mathbf{p}} t}\,
\end{equation} 
leads to the operator 
\begin{equation}\label{U_NSE}
    U_{\textrm{fSE}}^{(2)}
    =  \frac{i}{2(2\pi)^{4}}\int \frac{d \mathbf{p}}{E_{\mathbf{p}}^2} 
    b^\dagger(p\mu)b(p\mu) I_{f}(p),
\end{equation}
with the integral $I_{f}(p)$ for the electron self-energy
\begin{equation}\label{I_f_electron}
    I_{e^-}(p) =\int d^4 q
    \frac{2 e^2 P_{\alpha\beta}(p-q)(2p^\alpha q^\beta+g^{\alpha\beta}(m^2-pq))}{\left[ (p-q)^2-m_b^2+i\epsilon \right]
    \left[ q^2-m^2+i\epsilon \right]}
\end{equation}
and for the nucleon self-energy
\begin{multline}\label{I_f_nuclon}
    I_{N}(p) =\int d^4 q
    \frac{1}{\left[ (p-q)^2-m_b^2+i\epsilon \right]
    \left[ q^2-m^2+i\epsilon \right]}
    \\\times
    \bigg\{
        -2g^2 (3m^2-pq)
        -m_b^2\frac{f^2}{2m^2}(3m^2+pq)
        \\
        -6fg(p-q)^2
        + 4(\frac{g^2}{m_b^2}+\frac{f^2}{2m^2})
        (E_{\mathbf{p}}q_0 m_b^2
        \\
        +q_0^2\mathbf{p}^2+m^2\mathbf{q}^2
        -\mathbf{p}\mathbf{q}(q^2+m^2-\mathbf{p}\mathbf{q}))
    \bigg\}.
\end{multline}
Here $p=(E_{\mathbf{p}}, \mathbf{p})$, ${q=(q_0,\mathbf{q})}$, and $q_0 \neq E_{\mathbf{q}} = \sqrt{m^2+\mathbf{q}^2}$, 
whence the matrix elements
\begin{multline}
    \langle f|S_{\textrm{fSE}}^{(2)}|i \rangle = \frac{1}{2(2\pi)^{3}}
    \delta(E_f-E_i)\delta(\mathbf{p}'-\mathbf{p})\delta(\mathbf{k}'-\mathbf{k})
    \\\times
    \omega_{\mathbf{k}} \delta_{\mu'\mu}\delta_{\sigma'\sigma} I_{f}(p).
\end{multline}
Carefully separating the part of  the integral \eqref{I_f_nuclon} that cancels out with the contribution from $\langle f| V^{(2)}\,_{b^\dagger b} |i \rangle$ to the equation \eqref{ren_old_requirement_FD} we arrive to the nucleon mass shift
\begin{multline}\label{Nucl_FeynmanInt}
    \delta m_N^{(1\rho\textrm{-exc})} 
    = i \frac{1}{m(2\pi)^{4}}\int d^4 q
    \frac{1}{(p-q)^2-m_b^2+i\epsilon}
    \\ \times
    \frac{1}{q^2-m^2+i\epsilon}
    \bigg\{
        2g^2(2m^2 - pq) 
        + 3gf(p-q)^2
        \\+ 
        \frac{f^2}{4m^2}\big(
            m^2(3m^2+7q^2) - pq(6m^2+3m_b^2+4pq) 
        \big)
    \bigg\}.
\end{multline}
By using the equation \eqref{ren_old_requirement_FD} for the case of QED one can proof that the contribution to the integral \eqref{I_f_electron} from the first term in the definition \eqref{Pk}  leads to the momentum independent expression for the electron mass shift
\begin{multline}\label{FeynmanInt}
    \delta m_{e^-}^{(\textrm{1ph-exc})} 
    = i \frac{2e^2}{m(2\pi)^{4}}\int d^4 q
    \frac{1}{(p-q)^2-\lambda^2+i\epsilon}
    \\ \times
    \frac{2m^2 - pq}{q^2-m^2+i\epsilon},
\end{multline}
while the other terms in the definition of  $P_{\alpha\beta}(p-q)$ tensor lead to the cancellation of the matrix elements $\langle f| V^{(2)}\,_{b^\dagger b} |i \rangle \equiv \langle f| V_{\textrm{Coul}}\,_{b^\dagger b} |i \rangle$.
The integral \eqref{FeynmanInt} corresponds to the self-energy Feynman diagram Fig.~\ref{Feyn_self_energy_diagram}.
Now by integrating over $q_0$ in the r.h.s. of Eqs. \eqref{I_f_electron}-\eqref{I_f_nuclon} and taking into account the contributions from the simple poles $E_{\mathbf{q}}-i\epsilon$ and $E_{\mathbf{p}}+\omega_{\mathbf{p}-\mathbf{q}} -i\epsilon$, we get
\begin{multline}\label{Nucl_FeynmanInt_newform}
    \delta m_N^{(1\rho\textrm{-exc})} 
    = 
    \frac{1}{m(2\pi)^3}\int \frac{d \mathbf{q}}{E_{\mathbf{q}}}\frac{1}{(p-q)^2-m_b^2}
    \bigg\{
        g^2(2m^2\\-pq)
        +3gf(m^2-pq)+\frac{f^2}{4m^2}(m^2-pq)(5m^2-pq)
    \bigg\}
    \\+
    \frac{1}{2m(2\pi)^3}\int \frac{d \mathbf{k}}{\omega_{\mathbf{k}}}\frac{1}{(p+k)^2-m^2}
    \bigg\{
        2g^2(m^2-pk)\\
        +3gf m_b^2+f^2\left(m_b^2+\frac{(pk)^2}{2m^2}\right)
    \bigg\}
    = I_1 + I_2
\end{multline}
and
\begin{multline}\label{FeynmanInt_newform}
    \delta m_{e^-}^{(\textrm{1ph-exc})} 
    = \frac{e^2}{m(2\pi)^3}\int \frac{d \mathbf{q}}{E_{\mathbf{q}}}\frac{2m^2-pq}{(p-q)^2-\lambda^2}
    \\ +
    \frac{e^2}{m(2\pi)^3}\int \frac{d \mathbf{k}}{\omega_{\mathbf{k}}}\frac{m^2-pk}{(p+k)^2-m^2} = I'_1 + I'_2\,.
\end{multline}
\vspace{1mm}

\noindent
We see that the Dyson-Feynman formalism gives the same expressions for the mass shifts (\ref{delta_mN_fin})-(\ref{delta_mN_I2}) and (\ref{deltamUCT})-(\ref{delta_me-_I2}).
So we have found another proof of the momentum independence of the integrals (\ref{delta_mN_I1})-(\ref{delta_mN_I2}) and (\ref{delta_me-_I1})-(\ref{delta_me-_I2}).
But note that if we used the BPR and considered the matrix element $\langle a^\dagger \cdots \Omega_0|S(\alpha)|a^\dagger \cdots \Omega_0 \rangle$, we would arrive to the results which have the same analytical structure as $I_1$, $I_2$, $I'_1$, and $I'_2$ but with trial parameters $m_0$, $e_0$, $g_0$, and $f_0$ instead of the physical ones.

At this point, one should emphasize that similar steps when dealing with the $S$ operator become unnecessary if from the beginning we operate with interaction $K_I(\alpha_c)$ in the CPR of the Hamiltonian\footnote{To avoid confusion one needs to keep in mind that this relation refers to the case where functional $H$ depends on the set $\alpha$ or $\alpha_c$ (not a single argument).} $H(\alpha)=H(\alpha_c)$. 
This new form of $H$ 
being constructed via the sequential unitary transformations, gives a new unitarily equivalent form $S_\textrm{cloth}$ of the $S$ operator \cite{She2004}.

\section{Some elements of in(out) formalism. Interpolating fields\label{comparison_with_other}}

As well known, when evaluating the $S$--matrix in the $\rm{H}$ picture,
\begin{equation}
    S_{fi} = \langle f;out \mid  i;in \rangle
\end{equation}
one has to deal with the $in(out)$ states (see, e.g., \cite{GoldWat})
\begin{equation}
\left| \mathbf{k}_{1}\cdots\mathbf{k}_{n};\,in\,(out)\right\rangle
= a_{in\,(out)}^{^{\dagger }}\left( \mathbf{k}_{1}\right)
\,\cdots\,a_{in\,(out)}^{^{\dagger }}\left( \mathbf{k}_{n}\right) \left|\Omega\right\rangle
\label{eq_1}
\end{equation}
In particular, one-particle state:
\begin{equation}
\mid \mathbf{k};\,in\, (out) \left|\Omega\right\rangle = a_{in\,(out)}^{^{\dagger }}( \mathbf{k})
\left| \Omega\right\rangle,
\label{eq_2}
\end{equation}
where $ \left| \Omega \right\rangle $ is the physical vacuum.
The creation\,(destruction) $in(out)$ operators $a_{in(out)}^{^{\dagger }}$ \,
($a_{in(out)}$) meet the canonical commutation relations for bosons and fermions.

By definition, the $in(out)$ sates are eigenvectors of the energy-momentum operator $P^\mu=(H,\mathbf{P})$
\begin{equation}\label{inout_eigen}
   P^\mu| \mathbf k_1 \cdots \mathbf k_n;in(out)\rangle=(k_1^\mu+\cdots+k_n^\mu)| \mathbf k_1 \cdots \mathbf k_n;in(out)\rangle,
\end{equation}
with $k^\mu=(E_{\mathbf{k}},\mathbf{k})$.

For example, regarding an $in(out)$ state of a particle (say, electron or pion) and a bound
system $B$ (e.g., positronium or deuteron) one uses the ansatz
\begin{equation}
\left| \pi B;\,in (out)\right\rangle =a_{in\,(out)}^{^{\dagger }}\left( \mathbf{k}%
\right) \left| B\right\rangle ,  \label{eq_3}
\end{equation}
where $\left| B\right\rangle $ is the $H$ eigenvector.


First of all, we will show some common features of the CPR and the LSZ method in QFT with its creation (destruction) operators $a_{in\,(out)}^{\dagger }$ ($%
a_{in\,(out)}$). The latter enter the expansions in the plane waves
\begin{equation}
    f_{k}\left( x\right) =\left[ \left( 2\pi \right) ^{3}2k_{0}\right]
    ^{-1/2} e^{-ikx} ,\,\,k_{0}=\sqrt{\mathbf{k}^{2}+\mu ^{2}}
    \label{eq_5}
\end{equation} 
of the corresponding field operators $%
\varphi _{in(out)}\left( x\right) $ that meet the source-free Klein-Gordon (KG)
equation
\begin{equation}
\left( \Box _{x}+\mu ^{2}\right) \varphi _{in(out)}\left( x\right) =0.
\label{eq_4}
\end{equation}

For instance, in case of the {\it opposite-charged} scalar particles we
have
\begin{equation}
\varphi _{in}\left( x\right) =\int d \mathbf k\left[ A_{in}\left( \mathbf{k}\right)
f_{k}\left( x\right) +B_{in}^{\dagger }\left( \mathbf{k}\right) f_{k}^{*}\left(
x\right) \right],
\end{equation}
with the creation\,(destruction) $in$ operators that satisfy the canonical commutation relations for bosons and/or fermions.
Note also the properties
\begin{equation}
\left(f_{k^{\prime }}^{*}, f_{k} \right)
=\delta\left( \mathbf{k^{\prime }}-\mathbf{k}\right) ,\, \forall \,\{k^{\prime },\,k\},
\end{equation}
\begin{equation}
\left(f_{k^{\prime }}, f_{k} \right)
=0,\, \forall \,\{k^{\prime },\,k\} ,
\end{equation}
with the respect to the definition
\begin{equation}
    \left( f_{1},f_{2}\right) \equiv
        i\int d \mathbf x\left[ f_{1}\partial _{0}\,f_{2}-f_{2}\partial _{0}\,f_{1}\right]
\end{equation}
and the completeness condition
\begin{equation}
    \int d\mathbf{k}\, f_{k }^{*}\left( {x}\right) f_{k }\left(
{x}^{\prime }\right) |_{x_{0}=x_{0}^{\prime }}=\frac{1}{2k_{0}}\delta \left(
\mathbf{x}-\mathbf{x}^{\prime }\right).
\end{equation}
Using the relations we find
\begin{equation}
A_{in}\left( k\right) =\left( f_{k}^{*},\varphi_{in}\right) ,\,\,B_{in}^{\dagger }\left( k\right) =\left( \varphi _{in},f_{k}\right) .
\label{eq_11}
\end{equation}


Following a common practice we introduce the interpolating fields, these
mediators between the in\,(incoming)\,($ t \to -\infty $) fields and the out\,(outgoing)
($t \to +\infty $) fields.
Recall that for a given Heisenberg $\mathcal{H}$ field $\varphi({x})$ (for brevity, we restrict ourselves to a scalar field) the corresponding interpolating field is determined by
\begin{equation}
    \varphi _{int}\left( {x}\right) =\int d \mathbf k\left[ A_{int}\left( \mathbf{k},t\right)
    f_{k}\left( \mathbf{x}\right) +B_{int}^{\dagger }\left( \mathbf{k},t\right) f_{k}^{*}\left(
    \mathbf{x}\right) \right],
    \label{eq_33}
\end{equation}
where
\begin{equation}
    A_{int}\left( \mathbf{k}\right) =\left( f_{k}^{*},\varphi \right),\,\,
    B_{int}^{\dagger }\left( \mathbf{k}\right) =\left( \varphi ,f_{k}\right)
\end{equation}
and $f_{k}\left( \mathbf{x}\right) \equiv f_{k}\left( \mathbf{x}, t = 0\right)$.

At the same time, in accordance with conventional relation
\begin{equation}
    \varphi \left( {x}\right) =\exp \left( iHt\right) \varphi _{D}\left( \mathbf{x}%
    ,0\right) \exp \left( -iHt\right) ,  \label{eq_14}
\end{equation}
one can write
\begin{equation}
    \varphi \left( x\right) =\int d \mathbf k\left[ A\left( \mathbf{k},t\right) f_{k}\left(
    \mathbf{x},t\right) +B^{\dagger }\left( \mathbf{k},t\right) f_{k}^{*}\left( \mathbf{x}%
    ,t\right) \right].
    \label{eq_15}
\end{equation}

It is proved that
\begin{equation}
    \varphi _{int}\left( {x}\right) =\varphi \left( {x}\right).
    \label{eq_34}
\end{equation}

One should note the following links between $in(out)$ and clothed particle states,
viz., for the one interacting particle 
\begin{multline}
    \left| k;in(out)\right\rangle \equiv 
    A_{in(out)}^{^{\dagger }}\left( k\right) |\Omega\rangle 
    \\
    \equiv \lim_{t \to \mp\infty }
    A_{int}^{^{\dagger }}\left( k,t\right) |\Omega\rangle
    =A_{c}^{^{\dagger }}\left(k\right) |\Omega\rangle,
    \label{eq_46}
\end{multline}
To some extent, this relation does not seem unexpected, since both one-particle clothed state and $in(out)$ states, being equally normalized, are $H$ eigenvectors (cf. Eq.~\eqref{mass_def} and Eq.~\eqref{inout_eigen}). 
At the same time, for the two interacting particles we have
\begin{multline}
    \left| k_{1}k_{2};in(out)\right\rangle \equiv A_{in(out)}^{^{\dagger }}\left( k_{1}\right)A_{in(out)}^{^{\dagger }}\left( k_{2}\right) |\Omega\rangle 
    \\
    =\lim_{t \to \mp \infty } A_{int}^{^{\dagger }}\left(
    k_{1},t\right) A_{c}^{^{\dagger }}\left( k_{2}\right) |\Omega\rangle
    \\
    =\Omega_{c}^{(\mp)}A_{c}^{^{\dagger}}\left( k_{1}\right) A_{c}^{^{\dagger }}\left( k_{2}\right) |\Omega\rangle,
    \label{eq_50}
\end{multline}
with the Møller operators
$$
    \Omega_{c}^{(\mp)} \equiv \lim\limits_{t\to\mp\infty}
    \exp({iKt})\exp({-iK_Ft}).
$$
Furthermore, one should keep in mind the following recipe of practical calculations with
\begin{multline}
    \Omega^{ ( \pm ) } \mid E;c \rangle = \Omega^{ ( \pm ) }(E)\mid E;c \rangle \\
    = \pm i \lim_{\epsilon \to + 0} \epsilon
G(E \pm \imath \epsilon ) \mid E;c \rangle,
\label{eq_2.16}
\end{multline}
where we have introduced the notation $G(z) = (z - H)^{-1}$ for the Hamiltonian resolvent.
In its turn, the resolvent can be expressed through the corresponding $T$ or $R$
operator to make a path to known methods. 
Just such an approach has been used in papers \cite{SheDub2010, DubShe13} when studying the properties of two-nucleon systems.
Evidently, the appearance of these operators reflects the inclusion of the initial (final) state interaction effects into entrance (exit) reaction channels. In general, the interpolating fields are of great importance when
constructing the $S$-matrix that connects the distant past and the distant
future.



As to an illustration of the approach developed (cf. the example in Refs. \cite{ForSig2004,CatCia1984}) 
let us consider the electron-positron annihilation to the neutron-antineutron pair with the $S$-matrix
\begin{equation}
    S_{fi}(e^+ e^- \to n \bar{n})=\langle n \bar{n};out| e^+ e^-;in\rangle.
\end{equation}
Being immersed into calculations of the corresponding amplitude we apply the relation 
\begin{equation}
    S_{fi} = \langle \Psi_{f;c}^{(-)} \mid  \Psi_{i;c}^{(+)} \rangle
,\,\,\, \mid \Psi_{i(f);c}^{(\pm)} \rangle  = \Omega_{c}^{(\pm)} \mid i(f);c \rangle
\end{equation}
by introducing the respective distorted waves, if one uses the terminology adopted in the theory of nuclear reactions. We are retaining the term ``plane'' waves for the two-body clothed states
$b_c^\dag(n) d_c^\dag(\bar{n})| \Omega \rangle$
and
$b_c^\dag(e^-) d_c^\dag({e^+})| \Omega \rangle$.

In this context, we share the approach exposed in Ref. \cite{ForSig2004}: ``Rather than computing S matrix elements between usual states of the Fock space'' the authors construct ``dressed states that incorporate all long-range interactions''.

Finally, we would like to emphasize once more that within the CPR the mass counterterms in question are getting rid directly! in the Hamiltonian and can no longer appear in the $S$-matrix.

\section{Summary\label{summary}}

We have demonstrated how the mass shifts in the systems of interacting fermion and vector boson fields can be calculated within the CPR, where the total Hamiltonian and other generators of the Poincar\'e group take on a certain sparse structure in the Fock space.
We have chosen the two field models, viz., the interacting nucleon and $\rho$ meson fields along with the interacting electron and photon fields, to show how the UCT method allows us to get rid of the mass counterterms directly in the Hamiltonian.
Moreover, the contact terms that inevitably arise in the models with vector bosons \cite{Wein1} are cancelled too.
The respective mass counterterms are determined to compensate the other one-body terms that enter the Hamiltonian. 
These terms stem from the commutator $\frac{1}{2}\big[R^{(1)},V^{(1)}\big]$ and the operator $V^{(2)}$, where $V^{(1)}+V^{(2)}$ is the model interaction and $R^{(1)}$ is the generator of the corresponding clothing transformation.
Following the approach described above, the mass renormalization is done simultaneously with the construction of a new family of interactions between the physical particles (these quasiparticles in the CPR).
Explicit expressions for the interaction operators can be found in \cite{SheDub2010} (mesodynamics) and \cite{ArsKosShe2021} (QED). 

We have focused on deriving the mass shift in the second order in the coupling constants. But in general, the total mass shift is given by the series $\delta m=\delta m^{(2)}+\delta m^{(4)}+\cdots$. To evaluate the subsequent contributions to it one
needs to find the contributions from the more complicated commutators in the Campbell–Hausdorff expansion that has been employed when deriving the relationship \eqref{K_series}.
Doing so, we provide step by step the relation \eqref{mass_def}.

It turns out that the mass shifts obtained ($\delta m^{(2)}$) are covariant, that is not trivial because our path goes through the three-dimensional steps.
In particular, it means that in the local models the mass shifts values are independent of the particle momenta.
The experience acquired here has enabled, on the one hand, to reproduce the manifestly covariant results obtained within the Feynman techniques and, on the other hand, to derive a new representations \eqref{Nucl_FeynmanInt_newform} and \eqref{FeynmanInt_newform} for the one-loop Feynman integrals that can be depicted by the fermion self-energy diagram. 
In this context, one of us (A.S.) reminds of the comment by Walter Glöckle given during his visit to the Institut für Theoretische Physik, Ruhr-Universität Bochum: “Alex, our proof of the momentum independence of the nucleon mass shift (cf., integral (75) in Ref.~\cite{KruGlo1999}) took much more effort than yours \cite{KorShe2004}”.

\appendix*
\section{Derivation of the $NN \to NN$ and $N\bar{N} \to N\bar{N}$  interaction operators in the theory with vector coupling\label{KNN_derivation}}

To get the operators $K(NN \to NN)$ and $K(N\bar{N} \to N\bar{N})$ one needs to separate out the $b^\dagger b^\dagger b b$- and $b^\dagger d^\dagger b d$-type terms from the Hamiltonian \eqref{KI2}, respectively, i.e., from the commutator $\frac{1}{2}\big[R^{(1)},V^{(1)}\big]$ and the operator $V^{(2)}$. 

Let us consider the contribution from ${\big[R^{(1)},V^{(1)}\big]} = \big[ \mathcal{R}^{(1)} , \mathcal{V}^{(1)} \big] +
\big[ \mathcal{R}^{(1)} , \mathcal{V}^{(1) \dagger} \big] + \textrm{H.c.}$. According to Eqs. \eqref{V1_and_matrix}-\eqref{R1_and_matrix}, only the commutator $\big[ \mathcal{R}^{(1)} , \mathcal{V}^{(1) \dagger} \big]$ and its Hermitian conjugate contain terms of the required type.
One can rewrite $\mathcal{V}^{(1)}$ and $\mathcal{R}^{(1)}$ in a convenient form
\begin{multline*}
    \mathcal{V}^{(1)}= \frac{m}{\sqrt{2(2\pi)^3}}
    \int\frac{d \mathbf{k}}{\omega_{\mathbf{k}}} \frac{d \mathbf{p}_1}{E_{\mathbf{p}_1}} \frac{d \mathbf{p}_2}{E_{\mathbf{p}_2}}
    \delta(\mathbf{p}_1-\mathbf{p}_2-\mathbf{k})
    \\\times 
    e_\alpha(k\sigma) 
    \big\{
        v_{11}^\alpha(p_1\mu_1,p_2\mu_2,k)
        b^\dagger(p_1\mu_1) b(p_2\mu_2)
        \\+
        v_{12}^\alpha(p_1\mu_1,p_{2-}\mu_2,k)
        b^\dagger(p_1\mu_1) d^\dagger(p_{2-}\mu_2)
        \\+
        v_{21}^\alpha(p_{1-}\mu_1,p_2\mu_2,k)
        d(p_{1-}\mu_1) b(p_2\mu_2)
        \\-
        v_{22}^\alpha(p_{1-}\mu_1,p_{2-}\mu_2,k)
        d^\dagger(p_{2-}\mu_2) d(p_{1-}\mu_1) 
    \big\}
    a(k\sigma),
\end{multline*}
and
\begin{equation*}
    \mathcal{R}^{(1)}= \mathcal{V}^{(1)}
    \bigg(
        v_{\varepsilon' \varepsilon}^\alpha \mapsto
        \frac{v_{\varepsilon' \varepsilon}^\alpha}{(-1)^{\varepsilon'-1}E_{\mathbf{p}_1}-(-1)^{\varepsilon-1}E_{\mathbf{p}_2}-\omega_{\mathbf{k}}}
    \bigg),
\end{equation*}
where
\begin{align*}
    v_{11}^\alpha(p_1\mu_1,p_2\mu_2,k) = &
    \bar{u}(p_1\mu_1)
    (g\gamma^\alpha -\frac{f}{2m}ik_\beta\sigma^{\beta\alpha})
    u(p_2\mu_2) ,
    \nonumber
    \\
    v_{12}^\alpha(p_1\mu_1,p_2\mu_2,k) = &
    \bar{u}(p_1\mu_1)
    (g\gamma^\alpha -\frac{f}{2m}ik_\beta\sigma^{\beta\alpha})
    \upsilon(p_2\mu_2) ,
    \nonumber
    \\
    v_{21}^\alpha(p_1\mu_1,p_2\mu_2,k) = &
    \bar{\upsilon}(p_1\mu_1)
    (g\gamma^\alpha -\frac{f}{2m}ik_\beta\sigma^{\beta\alpha})
    u(p_2\mu_2) ,
    \nonumber
    \\
    v_{22}^\alpha(p_1\mu_1,p_2\mu_2,k) = &
    \bar{\upsilon}(p_1\mu_1)
    (g\gamma^\alpha -\frac{f}{2m}ik_\beta\sigma^{\beta\alpha})
    \upsilon(p_2\mu_2).
\end{align*}
Corresponding contributions from these operators are
\begin{multline}\label{rv_NN_to_NN}
    \big[ \mathcal{R}^{(1)} , \mathcal{V}^{(1) \dagger} \big]_{b^\dagger b^\dagger b b} =
    \frac{m^2}{(2\pi)^3} 
    \int d1' d2' d1 d2
    \\\times
    \frac{1}{2\omega_{\mathbf{k}}}
    \frac{f_1(1',2';1,2;k)}{E_{\mathbf{p}'_1}-E_{\mathbf{p}_1}-\omega_{\mathbf{k}}}
    \\\times
    \delta(\mathbf{p}'_1+\mathbf{p}'_2-\mathbf{p}_1-\mathbf{p}_2) 
    b^\dagger(1') b^\dagger(2') b(1) b(2),
\end{multline}
\vspace{-2mm}
\begin{multline}\label{rv_NaN_to_NaN}
    \big[ \mathcal{R}^{(1)} , \mathcal{V}^{(1) \dagger} \big]_{b^\dagger d^\dagger b d} =
    -\frac{m^2}{(2\pi)^3} 
    \int d1' d2' d1 d2
    \\\times
    \bigg[
        \frac{1}{2\omega_{\mathbf{k}}}
        \bigg\{
            \frac{f_2(1',2';1,2;k)}{E_{\mathbf{p}'_1}-E_{\mathbf{p}_1}-\omega_{\mathbf{k}}}
            +
            \frac{f_2^*(1,2;1',2';k_-)}{E_{\mathbf{p}'_2}-E_{\mathbf{p}_2}-\omega_{\mathbf{k}}}
        \bigg\}
        \\-
        \frac{1}{2\omega_{\mathbf{k}'}}
        \bigg\{
            \frac{f_3(1',2';1,2;k')}{E_{\mathbf{p}'_1}+E_{\mathbf{p}'_2}-\omega_{\mathbf{k}'}}
            +
            \frac{f_4(1',2';1,2;k'_-)}{-E_{\mathbf{p}_1}-E_{\mathbf{p}_2}-\omega_{\mathbf{k}'}}
        \bigg\}
    \bigg]
    \\\times
    \delta(\mathbf{p}'_1+\mathbf{p}'_2-\mathbf{p}_1-\mathbf{p}_2) 
    b^\dagger(1') d^\dagger(2') b(1) d(2),
\end{multline}
where $k=(\omega_{\mathbf{p}'_1-\mathbf{p}_1},\mathbf{p}'_1-\mathbf{p}_1)$, $k'=(\omega_{\mathbf{p}_1-\mathbf{p}_2},\mathbf{p}_1-\mathbf{p}_2)$,
\begin{align*}
    f_1(1',2';1,2;k)= &
    -P_{\alpha\beta}(k)
    v_{11}^\alpha(1',1,k)
    v_{11}^{\beta*}(2',2,k),
    \nonumber
    \\
    f_2(1',2';1,2;k)= &
    -P_{\alpha\beta}(k)
    v_{11}^\alpha(1',1,k)
    v_{22}^{\beta*}(2',2,k),
    \nonumber
    \\
    f_3(1',2';1,2;k)= &
    -P_{\alpha\beta}(k)
    v_{12}^\alpha(1',2',k)
    v_{12}^{\beta*}(1,2,k),
    \nonumber
    \\
    f_4(1',2';1,2;k)= &
    -P_{\alpha\beta}(k)
    v_{21}^{\alpha*}(2',1',k)
    v_{21}^\beta(2,1,k),
\end{align*}
and we denote for simplicity $\int d1'\equiv \sum\limits_{\mu'_1}\int\frac{d \mathbf{p}'_1}{E_{\mathbf{p}'_1}}$,\\ $(1') \equiv (\mathbf{p}'_1\mu'_1)$, etc.
Combining the contributions \eqref{rv_NN_to_NN}-\eqref{rv_NaN_to_NaN} with their Hermitian counterparts, we get

\begin{multline}
    \frac{1}{2}\big[R^{(1)},V^{(1)}\big]_{b^\dagger b^\dagger b b}=
    \frac{m^2}{2(2\pi)^3} 
    \int d1' d2' d1 d2
    \\\times
    \frac{1}{2\omega_{\mathbf{k}}} 
    \bigg\{ 
        \frac{f_1(1',2';1,2;k)}{E_{\mathbf{p}'_1}-E_{\mathbf{p}_1}-\omega_{\mathbf{k}}}
        +
        \frac{f_1(2',1';2,1;k)}{E_{\mathbf{p}_1}-E_{\mathbf{p}'_1}-\omega_{\mathbf{k}}}
    \bigg\}
    \\\times
    \delta(\mathbf{p}'_1+\mathbf{p}'_2-\mathbf{p}_1-\mathbf{p}_2) 
    b^\dagger(1') b^\dagger(2') b(1) b(2)
\end{multline}
and
\begin{multline}
    \frac{1}{2}\big[R^{(1)},V^{(1)}\big]_{b^\dagger d^\dagger b d} =
    -\frac{m^2}{2(2\pi)^3} 
    \int d1' d2' d1 d2
    \\\times
    \bigg[
        \frac{1}{2\omega_{\mathbf{k}}}
        \bigg\{
            \frac{f_2(1',2';1,2;k)}{E_{\mathbf{p}'_1}-E_{\mathbf{p}_1}-\omega_{\mathbf{k}}}
            +
            \frac{f_2(1',2';1,2;k)}{E_{\mathbf{p}_2}-E_{\mathbf{p}'_2}-\omega_{\mathbf{k}}}
            \\+
            \frac{f_2^*(1,2;1',2';k_-)}{E_{\mathbf{p}'_2}-E_{\mathbf{p}_2}-\omega_{\mathbf{k}}}
            +
            \frac{f_2^*(1,2;1',2';k_-)}{E_{\mathbf{p}_1}-E_{\mathbf{p}'_1}-\omega_{\mathbf{k}}}
        \bigg\}
        \\-
        \frac{1}{2\omega_{\mathbf{k}'}}
        \bigg\{
            \frac{f_3(1',2';1,2;k')}{E_{\mathbf{p}'_1}+E_{\mathbf{p}'_2}-\omega_{\mathbf{k}'}}
            +
            \frac{f_3(1',2';1,2;k')}{E_{\mathbf{p}_1}+E_{\mathbf{p}_2}-\omega_{\mathbf{k}'}}
            \\+
            \frac{f_4(1',2';1,2;k'_-)}{-E_{\mathbf{p}_1}-E_{\mathbf{p}_2}-\omega_{\mathbf{k}'}}
            +
            \frac{f_4(1',2';1,2;k'_-)}{-E_{\mathbf{p}'_1}-E_{\mathbf{p}'_2}-\omega_{\mathbf{k}'}}
        \bigg\}
    \bigg]
    \\\times
    \delta(\mathbf{p}'_1+\mathbf{p}'_2-\mathbf{p}_1-\mathbf{p}_2) 
    b^\dagger(1') d^\dagger(2') b(1) d(2).
\end{multline}
After the Gordon decompositions
\begin{equation}
    \begin{split}
        \bar{u}(1') i\sigma^{\alpha\beta}(p'_1-p_1)_\beta u(1) &=
        \bar{u}(1') \big( 2m \gamma^\alpha 
        \\&
        - (p'_1+p_1)^\alpha \big) u(1),
        \\
        \bar{u}(1') i\sigma^{\alpha\beta}(p'_1+p'_2)_\beta \upsilon(2') &=
        \bar{u}(1') \big( -2m \gamma^\alpha
        \\& 
        + (p'_1-p'_2)^\alpha \big) \upsilon(2'),
        \\
        \bar{\upsilon}(2) i\sigma^{\alpha\beta}(p_2-p'_2)_\beta \upsilon(2') &=
        \bar{\upsilon}(2) \big( - 2m \gamma^\alpha 
        \\&
        - (p_2+p'_2)^\alpha \big) \upsilon(2'),
    \end{split}
\end{equation}
we arrive to
\begin{widetext}
\begin{multline}\label{RVNN}
    \frac{1}{2}\big[R^{(1)},V^{(1)}\big]_{b^\dagger b^\dagger b b}=
    \frac{m^2}{2(2\pi)^3} 
    \int d1' d2' d1 d2\,
    \delta(\mathbf{p}'_1+\mathbf{p}'_2-\mathbf{p}_1-\mathbf{p}_2) 
    b^\dagger(1') b^\dagger(2') b(1) b(2)
    \\\times
    \Bigg\{
    \frac{1}{(p'_1-p_1)^2-m_b^2}
    \bar{u}(1') \big( (g+f)\gamma_\alpha - \frac{f}{2m}(p'_1+p_1)_\alpha \big) u(1)
    \bar{u}(2') \big( (g+f)\gamma^\alpha - \frac{f}{2m}(p'_2+p_2)^\alpha \big) u(2) 
    \\
    + (E_{\mathbf{p}'_1} + E_{\mathbf{p}'_2} - E_{\mathbf{p}_1} - E_{\mathbf{p}_2})
    \frac{f}{2m}
    \bar{u}(1') \frac{(g+f)\bm{\gamma} - \frac{f}{2m}(\mathbf{p}'_1+\mathbf{p}_1)}{(p'_1-p_1)^2-m_b^2} u(1)
    \bar{u}(2') \gamma^0\bm{\gamma} u(2)
    \\
    + \frac{g^2}{m_b^2} \bar{u}(1') \gamma_0 u(1) \bar{u}(2') \gamma_0 u(2)
    -\frac{f^2}{4m_b^2} \bar{u}(1') \gamma_0\bm{\gamma} u(1) \bar{u}(2') \gamma_0\bm{\gamma} u(2)
    \Bigg\}. 
\end{multline}
The last two terms in the curly brackets are contact, by definition.
At the same time we see that these terms are cancelled by those one stem from $V^{(2)}_{b^\dagger b^\dagger b b}$.
In such a way we come back to the result \eqref{KNN}.
Finally, a similar observation allows us to obtain the following expression for the $N\bar{N} \to N\bar{N}$ interaction between the clothed nucleons
\begin{multline}\label{KNaN}
    K(N\bar{N} \to N\bar{N})=
    \frac{m^2}{2(2\pi)^3} 
    \int d1' d2' d1 d2\,
    \delta(\mathbf{p}'_1+\mathbf{p}'_2-\mathbf{p}_1-\mathbf{p}_2) 
    b^\dagger(1') d^\dagger(2') b(1) d(2)
    \\\times
    \Bigg\{
        -\bigg[
            \frac{1}{(p'_1-p_1)^2-m_b^2} +
            \frac{1}{(p'_2-p_2)^2-m_b^2}
        \bigg]
        \bar{u}(1') \big( (g+f)\gamma_\alpha - \frac{f}{2m}(p'_1+p_1)_\alpha \big) u(1)
        \bar{\upsilon}(2) \big( (g+f)\gamma^\alpha + \frac{f}{2m}(p'_2+p_2)^\alpha \big) \upsilon(2') 
        \\
        - (E_{\mathbf{p}'_1} + E_{\mathbf{p}'_2} - E_{\mathbf{p}_1} - E_{\mathbf{p}_2})
        \frac{f}{2m}
        \bigg[
            \bar{u}(1') \frac{(g+f)\bm{\gamma} - \frac{f}{2m}(\mathbf{p}'_1+\mathbf{p}_1)}{(p'_1-p_1)^2-m_b^2} u(1)
            \bar{\upsilon}(2) \gamma^0\bm{\gamma} \upsilon(2')
            \\+
            \bar{u}(1') \gamma^0\bm{\gamma} u(1)
            \bar{\upsilon}(2) \frac{(g+f)\bm{\gamma} + \frac{f}{2m}(\mathbf{p}'_2+\mathbf{p}_2)}{(p'_2-p_2)^2-m_b^2} \upsilon(2')
        \bigg]
        +
        \big(
            E_{\mathbf{p}_2} \leftrightarrow -E_{\mathbf{p}'_1}, ~
            \mathbf{p}_2 \leftrightarrow -\mathbf{p}'_1, ~
            \textrm{and} ~
            \bar{\upsilon}(2) \leftrightarrow \bar{u}(1') 
        \big)
    \Bigg\}. 
\end{multline}
Here we confine ourselves to the application of the UCT method to $NN\to NN$ and $N\bar{N}\to N\bar{N}$ interaction operators.
Of course,
our calculations can be extended for deriving other interactions in the second order in the coupling constants, e.g., 
the annihilation $K(N\bar{N}\to \rho\rho)$ (see Ref.
\cite{ArsKosShe2021}).
\end{widetext}

The interactions \eqref{KNN} and \eqref{KNaN} together with $K(\bar{N}\bar{N} \rightarrow \bar{N}\bar{N})$ can be represented in the concise form 
\begin{multline}
    K_{NN} \equiv {K(NN \to NN)}
    + {K(N\bar{N} \to N\bar{N})} 
    \\
    + {K(\bar{N}\bar{N} \to \bar{N}\bar{N})},
\end{multline}
\begin{multline}
    K(NN \rightarrow NN)=
    \frac{m^2}{2(2\pi)^3}\int 
    d1' d2' d1 d2
    X_{11}(1',2',1,2)
    \\ \times
    \delta(\mathbf{p}'_1 + \mathbf{p}'_2 - \mathbf{p}_1 - \mathbf{p}_2)
    b^\dagger(1') b^\dagger(2') b(1) b(2),
\end{multline}
\begin{multline}
    K(N\bar{N} \rightarrow N\bar{N})=
    \frac{m^2}{2(2\pi)^3}\int 
    d1' d2' d1 d2
    \\ \times
    (X_{12}(1',2',1,2)
    +
    X_{21}(1',2',1,2))
    \\\times
    \delta(\mathbf{p}'_1 + \mathbf{p}'_2 - \mathbf{p}_1 - \mathbf{p}_2)
    b^\dagger(1') d^\dagger(2') b(1) d(2),
\end{multline}
\begin{multline}
    K(\bar{N}\bar{N} \rightarrow \bar{N}\bar{N})=
    \frac{m^2}{2(2\pi)^3}\int 
    d1' d2' d1 d2
    X_{22}(1',2',1,2)
    \\ \times
    \delta(\mathbf{p}'_1 + \mathbf{p}'_2 - \mathbf{p}_1 - \mathbf{p}_2)
    d^\dagger(1') d^\dagger(2') d(1) d(2).
\end{multline}
Such a form is convenient for more short representation of these interaction operators, viz.,
\begin{multline}\label{compact_KNN}
    K_{NN} 
    =
    -\frac{m^2}{2(2\pi)^3} \int 
    d1' d2' d1 d2
    \delta(\mathbf{p}'_1 + \mathbf{p}'_2 - \mathbf{p}_1 - \mathbf{p}_2)
    \\ \times
    :N^\dagger(1,1')
    X(1',2',1,2)
    N(2',2):,
\end{multline}
where one-fermion column $N$ and row $N^\dagger$ are composed of the pairs of the particle and antiparticle operators
\begin{equation}
    N(2',2)=
    \begin{bmatrix}
        b^\dagger(2')b(2) \\
        d^\dagger(2')d(2)
    \end{bmatrix}
\end{equation}
and the c-number matrix 
\begin{equation}\label{Xmatrix}
    X(1',2',1,2)=
    \begin{bmatrix}
        X_{11} & X_{12} \\
        X_{21} & X_{22}
    \end{bmatrix}(1',2',1,2).
\end{equation}

In the context of our nonlocal extension presented in Sec. \ref{Cutoff}, we would like to stress that such an extension ($K_{NN}^\textrm{nloc}$) leads to the replacement of the matrix \eqref{Xmatrix} by $X^\textrm{nloc}(1',2',1,2)$, which elements according to the rules
\begin{equation}
    \begin{split}
        &\hspace{15mm}
        \begin{matrix}
            \bar{u}(1)\Gamma u(2), &
            \bar{u}(1)\Gamma \upsilon(2),\\
            \bar{\upsilon}(1)\Gamma u(2), &
            \bar{\upsilon}(1)\Gamma \upsilon(2),
        \end{matrix}
        \\
        & \hspace{30mm} \downarrow
        \\
        &\begin{matrix}
            g_{11}(p_1,p_2) \bar{u}(1)\Gamma u(2), &
            g_{12}(p_1,p_2) \bar{u}(1)\Gamma \upsilon(2),\\
            g_{21}(p_1,p_2) \bar{\upsilon}(1)\Gamma u(2), &
            g_{22}(p_1,p_2) \bar{\upsilon}(1)\Gamma \upsilon(2),
        \end{matrix}
    \end{split}
\end{equation}
look like
\begin{multline}
        X^\textrm{nloc}_{\varepsilon' \varepsilon}(1',2',1,2)=
        g_{\varepsilon' \varepsilon'}(p'_1,p_1)g_{\varepsilon \varepsilon}(p'_2,p_2)
        \\
        \times X_{\varepsilon' \varepsilon}(1',2',1,2)
        \textrm{ for } \varepsilon' \leq\varepsilon 
\end{multline}
and
\begin{multline}
        X^\textrm{nloc}_{21}(1',2',1,2)=
        g_{12}(p'_1,p'_2)g_{21}(p_2,p_1)
        \\
        \times X_{21}(1',2',1,2).
\end{multline}
As mentioned above, these cutoff factors should be chosen to satisfy the $S$-matrix property \eqref{Smatrix_property}.
In accordance with a common practice  to calculate the $S$-matrix one starts with the Dyson-Feynman series \eqref{Sexp}. Following \cite{She2004} in the CPR we have an equivalent expansion
\begin{equation}
    S=\sum_{n=0}^\infty \frac{(-i)^n}{n!}
    \int d^4x_1 \dotsm \int d^4x_n P[K_I(x_1) \dotsm K_I(x_n)]
\end{equation}
where $K_I(x)=e^{iK_F t} K_I(\mathbf{x}) e^{-iK_F t}$ is an interaction density in the D-picture.
In our case the corresponding density can be written as
\begin{multline}
    K_{NN}(\mathbf{x})=
    -\frac{m^2}{2(2\pi)^6} \int 
    d1' d2' d1 d2
    \,e^{-i\mathbf{x}\cdot(\mathbf{p}'_1 + \mathbf{p}'_2 - \mathbf{p}_1 - \mathbf{p}_2)}
    \\ \times
    :N^\dagger(1,1') X(1',2',1,2) N(2',2):
\end{multline}
and taking into account the relations
\begin{equation}
    \begin{split}
        &e^{-iK_F t} b(p\mu) e^{iK_F t} =
        e^{i E_{\mathbf{p}} t} b(p\mu),
        \\
        &e^{-iK_F t} d(p\mu) e^{iK_F t} =
        e^{i E_{\mathbf{p}} t} d(p\mu)
    \end{split}
\end{equation}
we have
\begin{multline}
    K_{NN}(x)=
    -\frac{m^2}{2(2\pi)^6} \int 
    d1' d2' d1 d2
    \,e^{i x \cdot(p'_1 + p'_2 - p_1 - p_2)}
    \\ \times
    :N^\dagger(1,1') X(1',2',1,2) N(2',2):.
\end{multline}
Obviously, integrating $K_{NN}(x)$ over $x$ gives the delta function $\delta(p'_1 + p'_2 - p_1 - p_2)$. It automatically leads to the removal of the noncovariant terms in $X(1',2',1,2)$ because all of them have the factor $(E_{\mathbf{p}'_1} + E_{\mathbf{p}'_2} - E_{\mathbf{p}_1} - E_{\mathbf{p}_2})$ (see Eqs. \eqref{KNN} and \eqref{KNaN}). Thus, only explicitly covariant terms will enter the $S$ operator, that ensures $U_F(\Lambda) S U_F^{-1}(\Lambda)=S$ and the property \eqref{Smatrix_property}. 
At last, in the case of the nonlocal interaction $K_{NN}^\textrm{nloc}$, the latter property requires
$g_{\varepsilon' \varepsilon}(\Lambda p_1, \Lambda p_2)=g_{\varepsilon' \varepsilon}(p_1, p_2).$




%

\end{document}